\documentclass[11pt,draftcls,onecolumn]{IEEEtran}
\usepackage{stfloats}
\usepackage{subfig,epsfig} 
\usepackage{amsmath,amssymb,amsfonts}
\usepackage{multirow}
\usepackage{url} 
\usepackage{graphicx}
\graphicspath{{fig/}}

\newcommand{\diag}[1]{{\mbox{diag}\left\{{#1}\right\}}}

\begin{document}

\title{Robust Estimators for Variance-Based Device-Free Localization and Tracking}

\author{Yang Zhao and Neal Patwari\thanks{Y.~Zhao and N.~Patwari are with the Department
of Electrical and Computer Engineering, University of Utah, Salt Lake
City, USA. This material is based upon work supported by the National
Science Foundation under Grant Nos. \#0748206 and \#1035565.  E-mails:
yang.zhao@utah.edu and npatwari@ece.utah.edu.}}

\maketitle

\begin{abstract}
Human motion in the vicinity of a wireless link causes variations in the link received signal strength (RSS).
Device-free localization (DFL) systems, such as variance-based radio tomographic imaging (VRTI), use these RSS variations in a static wireless network to detect, locate and track people in the area of the network, even through walls. However, intrinsic motion, such as branches moving in the wind and rotating or vibrating machinery, also causes RSS variations which degrade the performance of a DFL system. 
In this paper, we propose and evaluate two estimators to reduce the impact of the variations caused by intrinsic motion. 
One estimator uses subspace decomposition, and the other estimator uses a least squares formulation. 
Experimental results show that both estimators reduce localization root mean squared error by about 40\% compared to VRTI. In addition, the Kalman filter tracking results from both estimators have 97\% of errors less than 1.3 m, more than 60\% improvement compared to tracking results from VRTI.
\end{abstract}

\section{Introduction} \label{S:Intro}

As an emerging technology, device-free localization (DFL) using radio frequency (RF) sensor networks has potential application in detecting intruders in industrial facilities, and helping police and firefighters track people inside a building during an emergency \cite{patwari11}.
In these scenarios, people to be located cannot be expected to participate in the localization system by carrying radio devices, thus standard radio localization techniques are not useful for these applications. 

Various RF measurements including ultra-wideband (UWB) and received signal strength (RSS) have been proposed and applied to detect, locate and track people who do not carry radio devices in an indoor environment \cite{chang04, song05, youssef07, wilson09c}. 
RSS measurements are inexpensive and available in standard wireless devices, and have been used in different DFL studies with surprising accuracy \cite{youssef07, zhang2007rf, wilson09c}.
These RSS-based DFL methods essentially use a windowed variance of RSS measured on static links.
For example, \cite{wilson09c} deploys an RF sensor network around a residential house and uses sample variance during a short window to track people walking inside the house; \cite{zhang2007rf} places RF sensors on the ceiling of a room, and track people using the RSSI dynamic, which is essentially the variance of RSS measurements, with and without people moving inside the room. 
In this paper we use windowed variance to describe the various functions of RSS measurements recently used in different DFL studies \cite{youssef07, zhang2007rf, wilson09c, viani2008object}, and we call these methods variance-based DFL methods. 

For variance-based DFL methods, variance can be caused by two types of motion: \emph{extrinsic motion} and \emph{intrinsic motion}. 
Extrinsic motion is defined as the motion of people and other objects that enter and leave the environment. 
Intrinsic motion is defined as the motion of objects that are intrinsic parts of the environment, objects which cannot be removed without fundamentally altering the environment. 
If a significant amount of windowed variance is caused by intrinsic motion, then it may be difficult to detect extrinsic motion. 
For example, rotating fans, leaves and branches swaying in wind, and moving or rotating machines in a factory all may impact the RSS measured on static links. Also, if RF sensors are vibrating or swaying in the wind, their RSS measurements change as a result. Even if the receiver moves by only a fraction of its wavelength, the RSS may vary by several orders of magnitude.
We call variance caused by intrinsic motion and extrinsic motion, the \emph{intrinsic signal} and \emph{extrinsic signal}, respectively.
We consider the intrinsic signal to be ``noise" because it does not relate to extrinsic motion which we wish to detect and track.

\begin{figure}[htbp]
  \centering
  \includegraphics[width=2.9in]{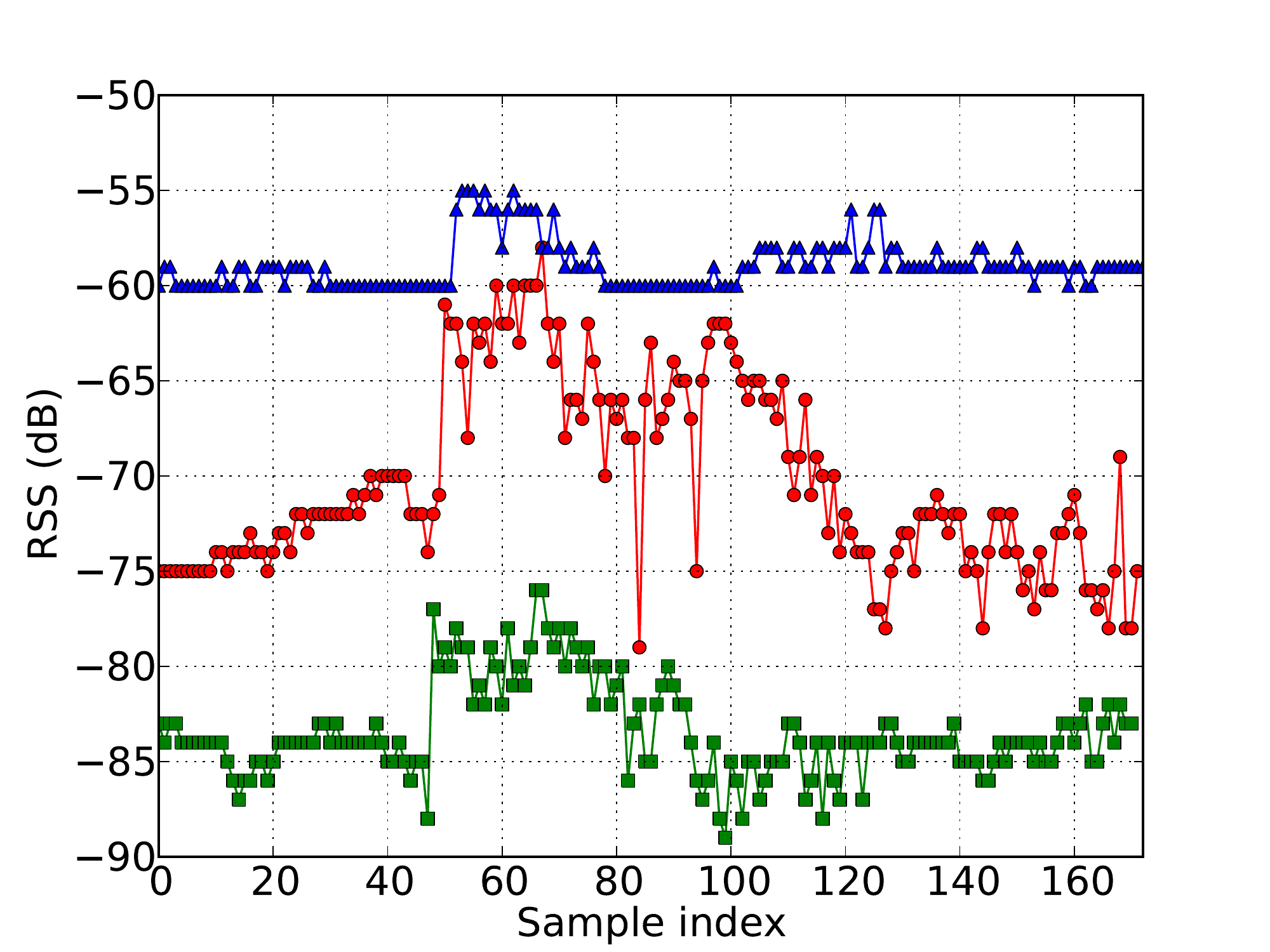}
  \caption{Intrinsic signal measurements: RSS measurements from three links during the calibration period when no people are present in the environment.}
  \label{F:linkPlot}
\end{figure}

This work is motivated by our inability to achieve the performance of 0.6~m average tracking error reported in \cite{wilson09c} in a repeat of the identical experiment in May, 2010. Our new experiment was performed at the same location and using the identical hardware, number of nodes, and software. Yet, in the new experiment, variance-based radio tomographic imaging (VRTI) does not always locate the person walking inside the house as accurately as reported in \cite{wilson09c}. Sometimes the position estimate error is as large as six meters, as shown in Figure~\ref{F:estimates_rti}.
Investigation of the experimental data quickly indicates the reason for the degradation: periods of high wind.
Consider the RSS measurements recorded during the calibration period, when no people are present inside the house. 
From the calibration measurements of \cite{wilson09c}, the standard deviations of RSS measurements are generally less than 2~dB. 
However, the RSS measurements from our May 2010 experiment are quite variable, as shown in Figure~\ref{F:linkPlot}. 
The RSS standard deviation can be up to 6~dB in a short time window. Considering there is no person moving inside the house, that is, no extrinsic motion during the calibration period, the high variations of RSS measurements must be caused by intrinsic motion, in this case, wind-induced motion.

The variance caused by intrinsic motion can affect both model-based DFL and fingerprint-based DFL methods. 
To apply various DFL methods in practical applications, the intrinsic signal needs to be identified and removed or reduced.
The subspace decomposition method has been used in spectral estimation, sensor array processing, and network anomaly detection \cite{stoica_moses, Schmidt86, Roy89, lakhina04}. We apply this method to VRTI, which leads to a new estimator we refer to as subspace variance-based radio tomography (SubVRT) \cite{zhao11secon}. 
Inspired by the fact that SubVRT makes use of the covariance matrix of link measurement and significantly reduces the impact of intrinsic motion, in this paper, we formulate a least squares (LS) solution \cite{Tarantola04} for VRTI which uses the inverse of the covariance matrix. We call this method least squares variance-based radio tomography (LSVRT).

The contribution of this paper is to propose and compare two estimators -- SubVRT and LSVRT to reduce the impact of intrinsic motion in DFL systems. 
Experimental results show that both estimators reduce the root mean squared error (RMSE) of the location estimate by more than 40\% compared to VRTI. 
Further, we use the Kalman filter to track people using localization estimates from SubVRT and LSVRT. The cumulative distribution functions (CDFs) of the tracking errors show that the tracking results from SubVRT have 97\% of errors less than 1.4~m, a 65\% improvement compared to VRTI, while 97\% of tracking errors from LSVRT are less than 1.2~m, a 70\% improvement. 

The rest of this paper is organized as follows: Section \ref{S:Methods} discusses the subspace decomposition method and least squares method for noise reduction in DFL. 
Section \ref{S:Experiments} describes the experiments, Section \ref{S:Results} shows the experimental results, and Section \ref{S:Tracking} investigates the Kalman filter tracking. Related work is presented in Section \ref{S:Related}, and the conclusion is given in Section \ref{S:Conclusion}.

\section{Methods} \label{S:Methods}

In this section, we formulate a variance-based DFL problem, introduce the subspace decomposition method, and propose our SubVRT estimator. After that, we use the measurement covariance matrix in a least squares (LS) formulation and propose another estimator, LSVRT. Finally, we discuss the connection between these two estimators.

\subsection{Problem statement} \label{S:Problem}

For an RF sensor network with $N$ sensors (radio transceivers) deployed at static locations, we use $\mathbf{z}_{s,j}$ to denote the coordinate of sensor $j$.
Each sensor makes an RSS measurement with many other sensors, and we use $s_{l,t}$ to denote the RSS measured at node $i_l$ sent by node $j_l$ at time $t$, where $i_l$ and $j_l$ are the receiver and transmitter number for link $l$, respectively. Time $t$ is discretized, thus $t \in \mathbb{Z}$.
We assume constant transmitter power so that changes in $s_{l,t}$ are due to the channel, not to the transmitter.
Then we denote the windowed RSS variance as:
\begin{equation}
  y_{l,t} = \frac{1}{m-1} \sum _{i=0} ^{m-1} (\bar{s}_{l,t} - s_{l,t-i} )^{2}
\end{equation}
where $m$ is the length of the window, and $\bar{s}_{l,t} = \frac{1}{m} \sum _{i=0} ^{m-1} s_{l,t-i}$ is the sample average in this window period.

Consider that the network has $L$ directional links on which we measure signal strength (in general, $L \le N(N-1)$).
We let $\mathbf{y}^{(t)} = [y_{1,t}, y_{2,t}, \cdots, y_{L,t}] ^{T}$ be the vector of windowed RSS variance from all $L$ links at time $t$. 
If we do not need to represent time, we simplify the notation to $\mathbf{y} = [y_{1}, y_{2}, \cdots, y_{L}] ^{T}$.
Then we use $\mathbf{y}_{c}$ to denote the calibration measurements collected during the calibration period, when no people are present in the environment; and we use $\mathbf{y}_{r}$ to denote the measurements from the real-time operation period.
The goal of DFL is to locate people during real-time operation. 

For VRTI, a model-based DFL method, the presence of human motion within $P$ voxels of a physical space is denoted by $\mathbf{x} = [x_1, x_2, ..., x_P]^{T}$, where $x_i=1$ if extrinsic motion occurs in voxel $i$, and $x_i=0$ otherwise. 
Work in \cite{wilson09c} has shown the efficacy of a linear model that relates the motion image $\mathbf{x}$ to the RSS variance $\mathbf{y}_r$:
\begin{equation} \label{E:model}
  \mathbf{y}_{r} = W \mathbf{x} + \mathbf{n}
\end{equation}
where $\mathbf{n}$ is an $ L \times 1$ noise vector including intrinsic motion and measurement noise, and $W$ is an $L \times P$ matrix representing the weighting of motion in each voxel on each link measurement.  
The weighting of voxel $p$ on link $l$ is formulated as \cite{wilson09c}:
\begin{equation} \label{E:W_lp}
   W_{l,p} = \frac{1}{\sqrt{d_{i_l, j_l}}}
  \begin{cases}
   \phi & \text{ if } d_{i_l,p} + d_{j_l,p} < d_{i_l, j_l} + \lambda \\
   0    & \text{ otherwise}
  \end{cases}
\end{equation}
where $d_{i_l,j_l}$ is the Euclidean distance between two sensors $i_{l}$, $j_{l}$ on link $l$ located at $\mathbf{z}_{s,i_{l}}$ and $\mathbf{z}_{s,j_{l}}$; $d_{j_l,p}$ is the Euclidean distance between sensor $j_{l}$ and $\mathbf{z}_p$, the center coordinate of voxel $p$; $d_{i_l,p}$ is the Euclidean distance between sensor $i_{l}$ and voxel $p$; $\lambda$ is a tunable parameter controlling the ellipse width, and $\phi$ is a constant scaling factor. 

Once we have the forward model, the localization problem becomes an inverse problem: to estimate $P$ dimensional position vector $\mathbf{x}$ from $L$ dimensional link measurement vector $\mathbf{y}_{r}$.
Certain regularization methods are necessary for this ill-posed inverse problem, and it is shown in \cite{wilson09c} that submeter localization accuracy can be achieved by using the Tikhonov regularization. 
Thus, we use the Tikhonov regularized VRTI solution, which is given as:
\begin{align} \label{E:Tik}
  \hat{\mathbf{x}} & = \Pi_{1} \mathbf{y}_{r} \nonumber \\
  \Pi_{1} & = (W^{T}W + \alpha Q^{T}Q)^{-1} W^{T} 
\end{align} 
where $Q$ is the Tikhonov matrix, and $\alpha$ is a regularization parameter.

\subsection{Subspace decomposition method} \label{S:Subspace} 

The subspace decomposition method has been widely used in spectral estimation, sensor array processing, etc. \cite{stoica_moses, lakhina04} to improve estimation performance in noise. It is closely related to principal component analysis (PCA), which is widely used in finding patterns in high dimensional data \cite{jolliffe02}.

From the $L$-dimensional calibration measurement vectors $\mathbf{y}_{c}$, we may estimate its covariance matrix $C_{\mathbf{y}_{c}}$ as:
\begin{equation} \label{E:C_y_c}
  C_{\mathbf{y}_c} = \frac{1}{M-1} \sum _{t=0} ^{M-1} (\mathbf{y}_{c}^{(t)} - \boldsymbol{\mu}_c) (\mathbf{y}_{c}^{(t)} - \boldsymbol{\mu}_c)^{T} 
\end{equation}
where $M$ is the number of sample measurements, $\mathbf{y}_{c}^{(t)}$ is the calibration measurement vector $\mathbf{y}_{c}$ at time $t$, $\boldsymbol{\mu}_{c} = \frac{1}{M} \sum _{t=0} ^{M-1} \mathbf{y}_{c}^{(t)}$ is the sample average.

Then, we perform singular value decomposition (SVD) on $C_{\mathbf{y}_{c}}$:
\begin{equation} \label{E:C_y}
  C_{\mathbf{y}_{c}} = U \Lambda U^{T}
\end{equation}
where the unitary matrix $U = [\mathbf{u}_{1} , \cdots , \mathbf{u}_{L}]$, and the diagonal matrix $\Lambda = \diag{\lambda_{1}, ..., \lambda_{L}}$.
Right multiplying $U$ on both sides of (\ref{E:C_y}), we have:
\begin{equation}
  C_{\mathbf{y}_{c}} \mathbf{u}_{i} = \lambda_{i} \mathbf{u}_{i}
  \label{E:eigen2}
\end{equation}
where $\mathbf{u}_{i}$ is the eigenvector corresponding to the eigenvalue $\lambda_{i}$.
If the eigenvalues are in descending order, the first principal component $\mathbf{u}_{1}$ points in the direction of the maximum variance in the calibration measurements, the second principal component $\mathbf{u}_{2}$ points in the direction of the maximum variance remaining in the measurements, and so on. 
If the first few eigenvalues are much larger than the others, then most of the variance in the calibration measurements can be captured by these principal components.

\begin{figure}[htbp]
  \centering
  \includegraphics[width=2.9in]{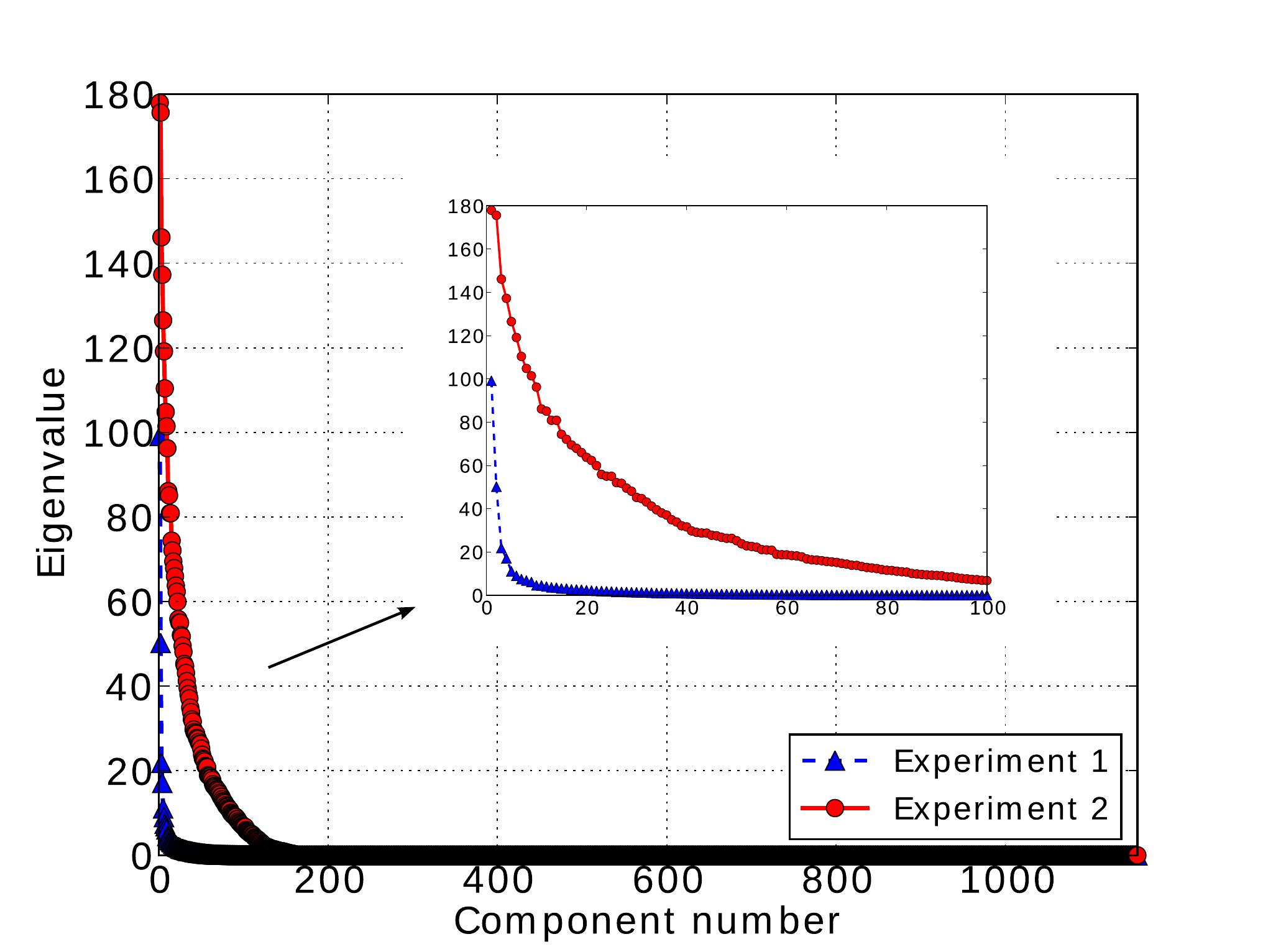}
  \caption{Scree plot.}
  \label{F:scree_plot}
\end{figure}

We perform the above PCA procedures on calibration measurements from two sets of experiments as described in Section~\ref{S:Experiments}. The eigenvalues of $C_{\mathbf{y}_c}$ from these experiments are shown in Figure~\ref{F:scree_plot}.
Because there is more intrinsic motion in Experiment~2, we see that the largest eigenvalue from Experiment~2 is almost twice as large as that from Experiment~1.
We also see that for Experiment~1, the first four eigenvalues are much larger than the other eigenvalues, thus the corresponding eigenvectors can capture most of the variation in the measurements. 
However, for Experiment~2, there are more large-valued eigenvalues, and more eigenvectors are necessary to represent the major variation in the measurements. 

From the scree plot, we decide how many principal components, $k$, are necessary to capture the majority of the variations (we discuss selection of $k$ in more detail in Section~\ref{S:Results_dis}). Then, we use a lower dimensional space spanned by these principal components to represent the space containing the majority of the intrinsic signal measurements. Thus, in subspace decomposition, we divide all the principal components into two sets: $\hat{U}=[\mathbf{u}_{1}, \mathbf{u}_{2}, \cdots, \mathbf{u}_{k} ]$ and $\tilde{U} = [\mathbf{u}_{k+1}, \mathbf{u}_{k+2}, \cdots, \mathbf{u}_{L} ]$.
Since the variance during the calibration period is caused by intrinsic motion, that is, the variance captured by $\hat{U}$ is intrinsic signal, we call the subspace spanned by $\hat{U}$ the intrinsic subspace, and the other subspace spanned by $\tilde{U}$ the extrinsic subspace. Once the two subspaces are constructed, we can decompose the measurement vector $\mathbf{y}$ into two components -- intrinsic signal component $\hat{\mathbf{y}}$ and extrinsic signal component $\tilde{\mathbf{y}}$:
\begin{equation}
  \mathbf{y} = \hat{\mathbf{y}} + \tilde{\mathbf{y}}.
\end{equation}
Since the principal components are orthogonal, the intrinsic signal component $\hat{\mathbf{y}}$ and the extrinsic signal component $\tilde{\mathbf{y}}$ can be formed by projecting $\mathbf{y}$ onto the intrinsic subspace and the extrinsic subspace, respectively:
\begin{align}
  \hat{\mathbf{y}} & = \Pi _{I} \mathbf{y} = \hat{U} \hat{U}^{T} \mathbf{y} \\
  \tilde{\mathbf{y}} & = \Pi _{E} \mathbf{y} = (I - \hat{U} \hat{U}^{T}) \mathbf{y} \label{E:tilde_y}
\end{align}
where $\Pi_{I}= \hat{U} \hat{U}^{T}$ is the projection matrix for the intrinsic subspace, and $\Pi_{E}=I-\Pi_{I}$ is the projection matrix for the extrinsic subspace.

The key idea of SubVRT is to use the decomposed extrinsic signal component of the measurements in VRTI.
We project the real-time measurement vector $\mathbf{y}_{r}$ onto the extrinsic subspace to obtain the extrinsic signal component $\tilde{\mathbf{y}}_{r} = (I - \hat{U} \hat{U}^{T})\mathbf{y}_{r}$. 
Then, we replace $\mathbf{y}_{r}$ in (\ref{E:Tik}) with $\tilde{\mathbf{y}}_{r}$ and obtain the solution of SubVRT:
\begin{align} \label{E:SSD-VRTI}
  \hat{\mathbf{x}}_{Sub} & = \Pi_{2} \mathbf{y}_r \nonumber \\
  \Pi_{2} & = (W^{T} W + \alpha Q^{T} Q) ^{-1} W^{T} (I-\hat{U}\hat{U}^{T}).
\end{align}
From (\ref{E:SSD-VRTI}), we see that the solution is a linear transformation of the measurement vector. The transformation matrix $\Pi_{2}$  is the product of the transformation matrix $\Pi_{1}$ in (\ref{E:Tik}) with the projection matrix for the extrinsic subspace $\Pi_{E}$: $\Pi_{2} = \Pi_{1} \Pi_{E}$.
Since the transformation matrix $\Pi_{2}$ does not depend on instantaneous real-time measurements, it can be pre-calculated, and it is easy to implement SubVRT for real-time applications. 
Calculation of $\hat{\mathbf{x}}$ from $\mathbf{y}_{r}$ requires $LP$ multiplications and additions.

\subsection{Least squares method}
SubVRT performs SVD on the calibration measurement covariance matrix. Here, we introduce our LSVRT estimator formulated as a least squares (LS) solution, which uses the inverse of the covariance matrix.

\subsubsection{Formulation}

To derive the least squares solution to the linear model expressed in (\ref{E:model}), the cost function can be written as \cite{Tarantola04}:
\begin{align} \label{E:misfit}
  J(\mathbf{x}) & = \| W \mathbf{x} - \mathbf{y}_{r} \| _{C_{\mathbf{n}}} ^{2} + \| \mathbf{x} - \mathbf{x}_{a} \| _{C_{\mathbf{x}}} ^{2} \\ \nonumber
  & = (\mathbf{y}_{r} - W\mathbf{x})^{T} C_{\mathbf{n}}^{-1} (\mathbf{y}_{r} - W\mathbf{x}) + (\mathbf{x} - \mathbf{x}_{a})^{T} C_{\mathbf{x}}^{-1} (\mathbf{x} - \mathbf{x}_{a})
\end{align}
where $\| \mathbf{n} \|_{C_{\mathbf{n}}} ^{2}$ indicates weighted quadratic distance $\mathbf{n}^{T} C_{\mathbf{n}} ^{-1} \mathbf{n} $, 
$C_{\mathbf{n}}$ is the covariance matrix of $\mathbf{n}$, 
$\mathbf{x}_{a}$ is the prior mean of $\mathbf{x}$, and
$C_{\mathbf{x}}$ is the covariance matrix of $\mathbf{x}$.

Taking the derivative of (\ref{E:misfit}) and setting it to zero results in the LSVRT solution:
\begin{equation}  \label{E:x_hat1}
  \hat{\mathbf{x}}_{LS} = (W^{T}C_{\mathbf{n}}^{-1}W + C_{\mathbf{x}}^{-1}) ^{-1} (W^{T} C_{\mathbf{n}} ^{-1} \mathbf{y}_{r} + C_{\mathbf{x}}^{-1} \mathbf{x}_{a}).
\end{equation}
Since the prior information $\mathbf{x}_{a}$ can be included in the tracking period, here we assume $\mathbf{x}_{a}$ is zero, then (\ref{E:x_hat1}) becomes:
\begin{align} \label{E:x_hat}
  \hat{\mathbf{x}}_{LS} & = \Pi_{3} \mathbf{y}_{r} \nonumber \\
  \Pi_{3} & = (W^{T} C_{\mathbf{n}}^{-1}W + C_{\mathbf{x}}^{-1}) ^{-1} W^{T} C_{\mathbf{n}} ^{-1}.
\end{align} 
The LSVRT formulation can be also justified from a Bayesian perspective. 
If we assume $\mathbf{y}_{r}$ conditioned on $\mathbf{x}$ is Gaussian distributed with mean $W\mathbf{x}$ and covariance matrix $C_{\mathbf{n}}$, and $\mathbf{x}$ is Gaussian distributed with mean $\mathbf{x}_{a}$ and covariance matrix $C_{\mathbf{x}}$, then maximizing the posteriori distribution $p(\mathbf{x} \vert \mathbf{y}_{r})$ is equivalent to minimizing the cost function in (\ref{E:misfit}), thus the maximum a posteriori (MAP) solution is the same as (\ref{E:x_hat1}).

\subsubsection{Covariance matrix $C_{\mathbf{n}}$}

From the LSVRT solution (\ref{E:x_hat1}), we see that the inverse of the covariance matrix $C_{\mathbf{n}}$ (a.k.a., the precision matrix) is needed. 
We may use the sample covariance matrix if the sample size $M$ is greater than the number of link measurements $L$. However, for an RF sensor network with $L$ directional links, $M$ is typically less than $L$. Thus, for high dimensional problems, the sample covariance matrix becomes an ill-posed estimator, it cannot be inverted to compute the precision matrix.

For high dimensional covariance matrix estimation problems, many types of regularized covariance matrix estimators have been proposed \cite{Ledoit04, Chen10}. Here, we use the Ledoit-Wolf estimator, which is a linear combination of the sample covariance matrix and a scaled identity matrix, and is asymptotically optimal for any distribution \cite{Ledoit04}:
\begin{equation} \label{E:LW}
  C_{\mathbf{n}} = \nu \mu I + (1-\nu) C_{\mathbf{n}}^{*}
\end{equation}
where $C_{\mathbf{n}}^{*}$ is the sample covariance matrix, $\mu$ is the scaling parameter for the identity matrix $I$, and $\nu$ is the shrinkage parameter that shrinks the sample covariance towards the scaled identity matrix. 
Since there is no extrinsic motion during calibration period, that is, $\mathbf{x} = 0$, thus $\mathbf{y}_{c} = \mathbf{n}$, and we approximate $C_{\mathbf{n}}^{*} = C_{\mathbf{y}_{c}}$.
Then we follow \cite{Ledoit04} to calculate parameters $\nu$ and $\mu$. 
From the Bayesian perspective, this covariance matrix estimator can be seen as the combination of the prior information and sample information of the covariance matrix.

\subsubsection{Covariance matrix $C_{\mathbf{x}}$}
The LSVRT solution also requires the covariance matrix $C_{\mathbf{x}}$. 
As a means to generate a general statistical model for $C_{\mathbf{x}}$, we assume that the positions of people in the environment can be modeled as a Poisson process.
Poisson processes are commonly used for modeling the distribution of randomly arranged points in space. 

Analysis of Poisson point processes leads to a covariance function that is approximately exponentially decaying \cite{agrawal09}, and the exponential spatial covariance model is shown to be effective to locate people in an RF sensor network \cite{wilson09a}.
Thus, in this paper, we use an exponentially-decaying function as the covariance matrix of the human motion.
\begin{equation} \label{E:exp_func}
  C_{\mathbf{x}} = \frac{\sigma_{x}^{2}}{\delta} \exp \left( - \frac{\Vert \mathbf{x}_{j} - \mathbf{x}_{i} \Vert _{l_{2}}}{\delta} \right)
\end{equation}
where $\sigma_{x}^{2}$ is the variance of the human motion, $\delta$ is a space constant, and $\Vert \mathbf{x}_{j} - \mathbf{x}_{i} \Vert _{l_{2}}$ is the Euclidian distance between $\mathbf{x}_i$ and $\mathbf{x}_j$.

\subsection{Discussion}

The SubVRT estimator and the LSVRT estimator are closely related. 
LSVRT needs to calculate the inverse of the covariance matrix $C_{\mathbf{n}}$, while SubVRT needs to perform SVD on the sample covariance matrix $C_{\mathbf{y}_{c}}$. In this section, we show connections between these two estimators.

First, for SubVRT, once we choose the parameter $k$, we can find a diagonal matrix $S=\diag{\underbrace{0,0,\cdots, 0,}_{k} 1, 1, \cdots, 1 }$ such that $U S U^{T} = I-\hat{U} \hat{U}^{T}$. Then, the project matrix for the SubVRT solution can be rewriten as:
\begin{equation} \label{E:Pi2}
  \Pi_{2} = (W^{T} W + \alpha Q^{T} Q) ^{-1} W^{T} U S U^{T}.
\end{equation}

For the LSVRT solution (\ref{E:x_hat}) and the Ledoit-Wolf covariance estimator in (\ref{E:LW}), if we approximate $C_{\mathbf{n}}^{*} = C_{\mathbf{y}_{c}}$, then the inverse of $C_{\mathbf{n}}$ can be written as:
\begin{equation} \label{E:C_y_inv}
  C_{\mathbf{n}} ^{-1} = \frac{1}{\nu \mu} I + \frac{1}{1- \nu} C_{\mathbf{y}_{c}}^{-1}.
\end{equation}
Substituting (\ref{E:C_y}) in (\ref{E:C_y_inv}), we express $C_{\mathbf{n}}^{-1}$ in terms of $\Lambda$:
\begin{equation} \label{E:C_y_gamma}
  C_{\mathbf{n}} ^{-1} =  U c_{1} (\Lambda ^{-1} + c_{2}I) U^{T}
\end{equation}
where $c_{1} = \frac{1}{1-\nu}$, and $c_{2} = \frac{1-\nu}{\nu \mu}$.
Replacing the second $C_{\mathbf{n}} ^{-1}$ in (\ref{E:x_hat}) by (\ref{E:C_y_gamma}), the project matrix for the LSVRT solution becomes:
\begin{equation} \label{E:Pi3}
  \Pi_{3} = (W^{T} C_{\mathbf{n}} ^{-1} W + C_{\mathbf{x}}^{-1} ) ^{-1} W^{T} U c_{1} (\Lambda ^{-1} + c_2 I) U^{T}.
\end{equation}

Now we compare the two projection matrices (\ref{E:Pi2}) and (\ref{E:Pi3}) in the SubVRT and LSVRT solutions.
From the latter part of (\ref{E:Pi3}), we see that LSVRT uses $c_1(\Lambda ^{-1} + c_2 I)$ to give less weights to the linear combinations of measurements in the eigen-space with high variance (large eigenvalues).
For SubVRT, the diagonal matrix $S$ in (\ref{E:Pi2}) is used to directly remove eigenvectors that correspond to the first $k$ largest eigenvalues.
From the former part of (\ref{E:Pi2}) and (\ref{E:Pi3}), we see that the inverse of the covariance matrix $C_{\mathbf{x}}^{-1}$ in the LSVRT solution plays the same role of regularization as the term $\alpha Q^{T} Q$ in the SubVRT solution. 
We also see that the LSVRT estimator includes the precision matrix $C_{\mathbf{n}} ^{-1}$ as a weight matrix in $W^{T} C_{\mathbf{n}} ^{-1} W$, while the SubVRT estimator just uses $W^{T}W$.

\section{Experiments} \label{S:Experiments}

We use measurements from two sets of experiments in this paper. 
We use the data set from the measurements conducted in March, 2009 reported by \cite{wilson09c}. We call this data set Experiment~1. The second experiment is a new experiment performed in May, 2010 at the same residential house, which we call Experiment~2. 
In both experiments, thirty-four TelosB nodes are deployed outside the living room of the house. 
As shown in Figure~\ref{F:layout}, eight nodes are placed on the table in the kitchen, six nodes are placed on boards extended outside the windows of the living room. 
The other twenty nodes are all placed on polyvinyl chloride (PVC) stands outside the house.
All thirty-four nodes are programmed with TinyOS program Spin \cite{Spin}, and a basestation connected to a laptop is used to collect pairwise RSS measurements from these nodes. 

\begin{figure*}[htb]
  \centering
  \subfloat[Experiment 1]{\label{F:exp1}\includegraphics[width=0.38\textwidth, height=0.24\textwidth]{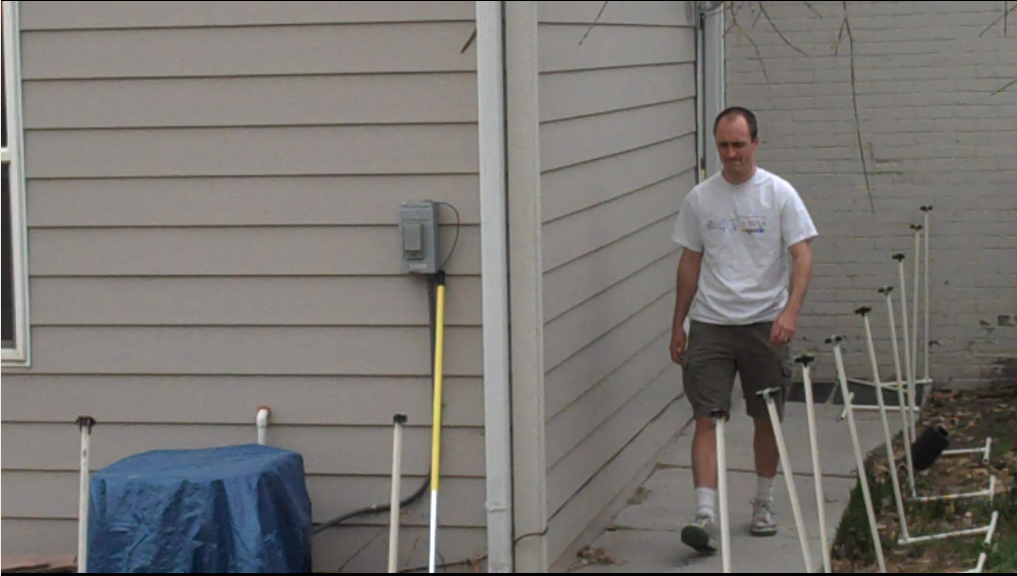}}$\,$
  \subfloat[Experiment 2]{\label{F:exp2}\includegraphics[width=0.38\textwidth, height=0.24\textwidth]{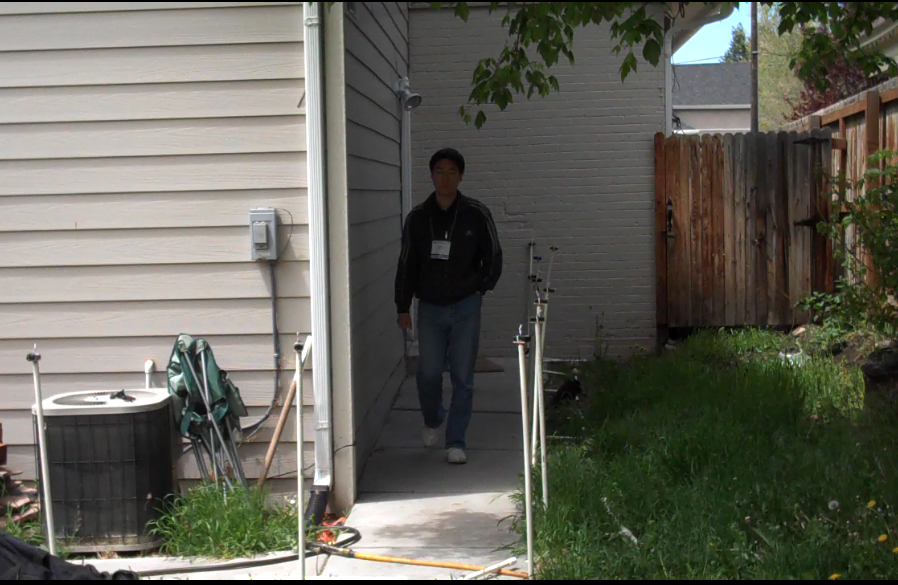}}
  \caption{Pictures of two experiments.}
\end{figure*}

\begin{figure}[htbp]
  \centering
  \includegraphics[width=2.9in]{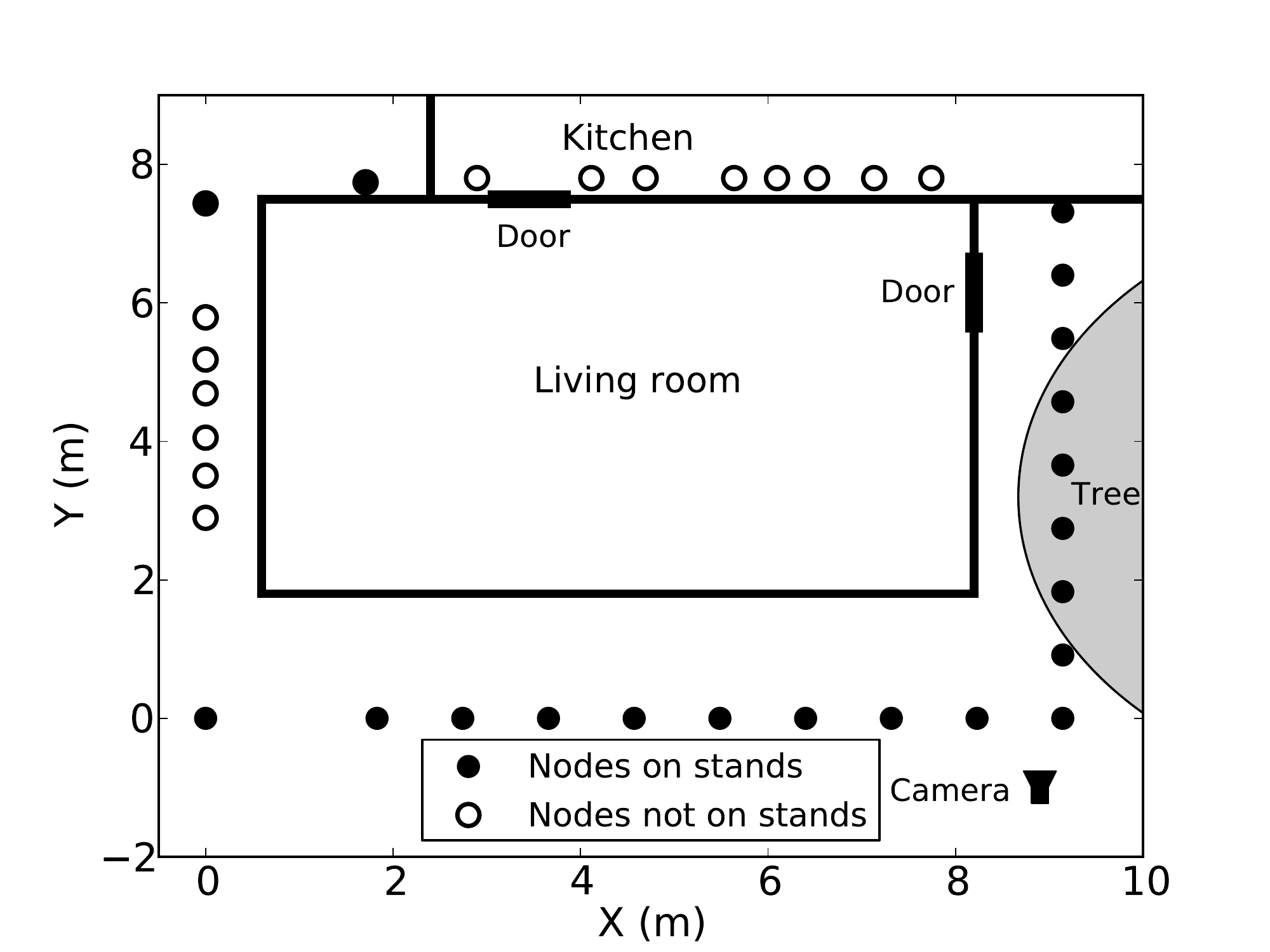}
  \caption{Experimental layout of Experiment~2. The shade area is covered by tree branches and leaves.}
  \label{F:layout}
\end{figure}

Both experiments are performed using the following procedure. 
Before people start to walk in the living room, a calibration is performed with no people (no extrinsic motion) in the experimental area. 
The duration of the calibration period of Experiment~1 is about 47 seconds, and $M = 140$ measurements are recorded for each link; while for Experiment~2, $M=170$ measurements are recorded for each link during a 57 second calibration period. Compared to $L=1122$ directional links, $M$ is much smaller than $L$.
Next, a person walks around a marked path in the living room at a constant speed, using a metronome and a metered path so that the position of the person at any particular time is known.

These two through-wall experiments use the same hardware and software, and are performed following the same procedure.
However, the main difference between these two experiments is the season.
Experiment~1 is performed on a clear winter day, while Experiment~2 is performed on a windy day in late spring. As shown in Figure~\ref{F:exp1}, there are no leaves on branches and no wind is observed from the video of Experiment~1. However, from the video recorded during Experiment~2 (one snapshot is shown in Figure~\ref{F:exp2}), we observe that wind causes grass, leaves and branches to sway \cite{SPAN_RTI}. The wind also causes the PVC stands supporting the nodes to move. 
The swaying of leaves and branches and the movement of the PVC stands are intrinsic parts of the environment, which cannot be avoided, even when no people are present in the environment. Thus, the difference between Experiments~1 and 2 is that Experiment~2 has more intrinsic motion.

\section{Results} \label{S:Results}

\subsection{Eigen-network results}

As described in Section \ref{S:Subspace}, each of the principal components used to construct the intrinsic subspace is an eigenvector of the covariance matrix of the network measurements, and each element in an eigenvector is from an individual link, we refer these eigenvectors $\mathbf{u}_{i}$ as ``eigen-networks". 

\begin{figure}[htbp]
  \centering
  \includegraphics[width=2.9in]{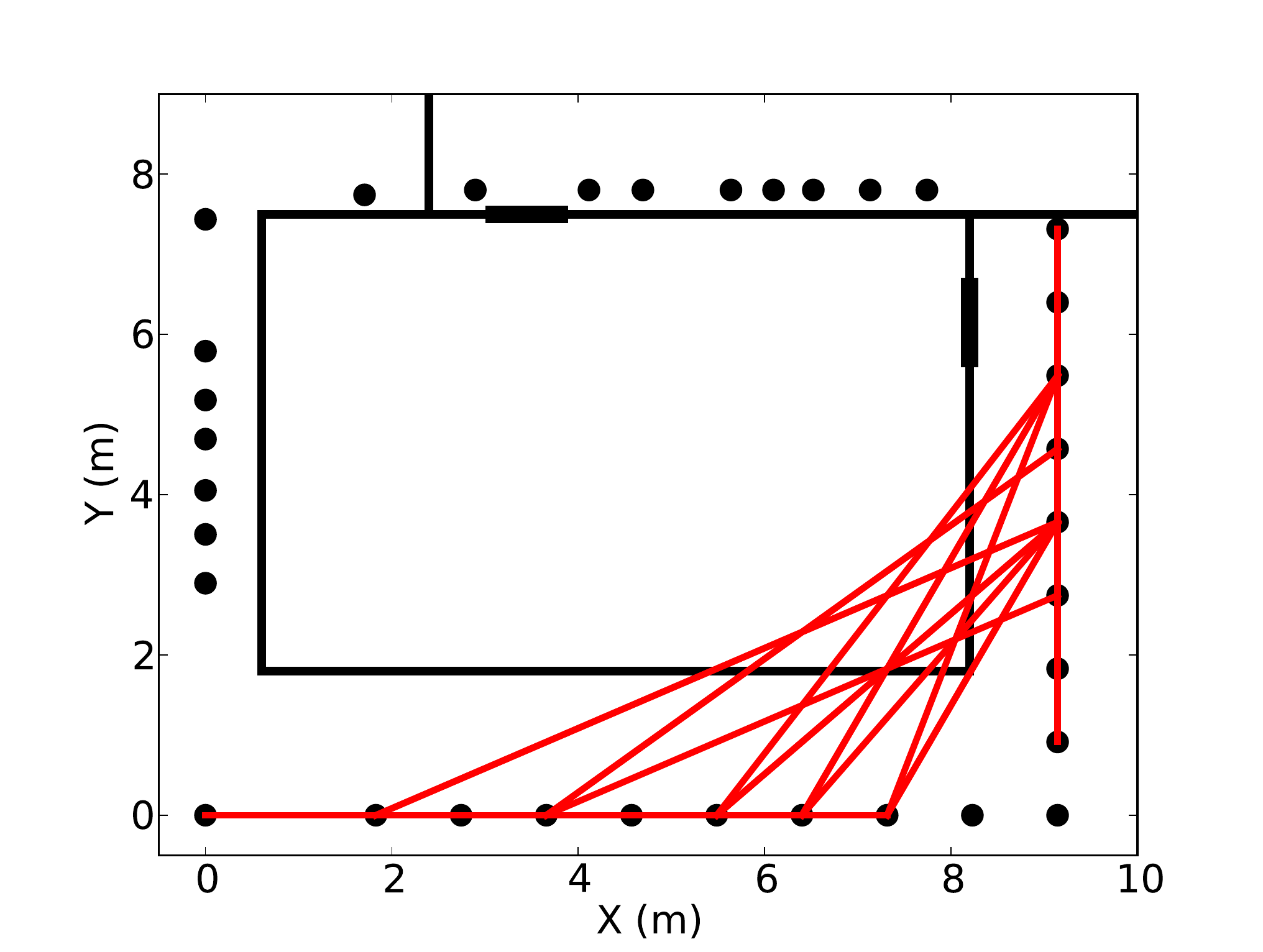}
  \caption{First eigen-network: Links with $u_{1l} > 30\%$ of $\operatorname*{max}_{l} u_{1l}$.} 
  \label{F:noise_source}
\end{figure}

Since the first eigen-network $\mathbf{u}_{1}= [u_{11}, u_{12}, \cdots, u_{1L}]^{T}$ points in the direction of the maximum variance of the calibration measurements $\mathbf{y}_c$, we show the first eigen-network $\mathbf{u}_{1}$ graphically in Figure~\ref{F:noise_source}.
We see the links with $u_{1l}$ values higher than $30\%$ of the maximum value are all in the lower right side of the house.
This is consistent with our observation that the intrinsic motion of the leaves and branches on the tree located to the right side of the house causes significant variations in the RSS measured on links likely to have RF propagation through the branches and leaves. Note that links with high $u_{1l}$ values all have at least one end point near the tree. In particular, links which are likely to see significant diffraction around the bottom-right corner of the house have high $u_{1l}$ values. The leaves and branches almost touch this corner, as seen in Figure~\ref{F:exp2}.
Not only do these links measure high RSS variance during the calibration period, they do so simultaneously. That is, the fact that these links have high positive $u_{1l}$ values indicates that when one of these links experiences increased RSS variance, the other links also measure increased RSS variance. Thus, the first eigen-network $\mathbf{u}_{1}$ becomes a spatial signature  for intrinsic motion-induced RSS variance.
When we see this linear combination in $\mathbf{y}_r$, we should attribute it to intrinsic, rather than extrinsic motion.
These observations about the source of RSS variance on links support the idea that intrinsic motion in the environment causes increased RSS variance simultaneously on multiple links.

\subsection{Localization results} \label{S:loc_results}

Now, we evaluate VRTI, SubVRT and LSVRT using measurements from Experiments~1 and 2.
From these three estimators, we obtain reconstructed motion images, and the position of the moving person can be estimated by finding the center coordinate of the voxel with maximum value. 
Specifically, a localization estimate is defined as: 
\[
  \hat{\mathbf{z}} = \mathbf{z}_q \quad \mbox{where} \quad q = \operatorname*{arg\,max}_{p} \hat{x}_p
\]
where $\mathbf{z}_q$ is the center coordinate of voxel $q$, and $\hat{x}_p$ is the $p$th element of the estimate $\hat{\mathbf{x}} = [\hat{x}_1, \hat{x}_2, ..., \hat{x}_P]^{T}$ from (\ref{E:Tik}), (\ref{E:SSD-VRTI}) or (\ref{E:x_hat}).
Then, the localization error is defined as: $ e_{loc} = \| \hat{\mathbf{z}} - \mathbf{z} \| _{l_2}$, where $\mathbf{z}$ is the actual position of the person, and $l_2$ indicates the Euclidean norm.

\begin{figure}[htbp]
  \centering
  \includegraphics[width=3.1in]{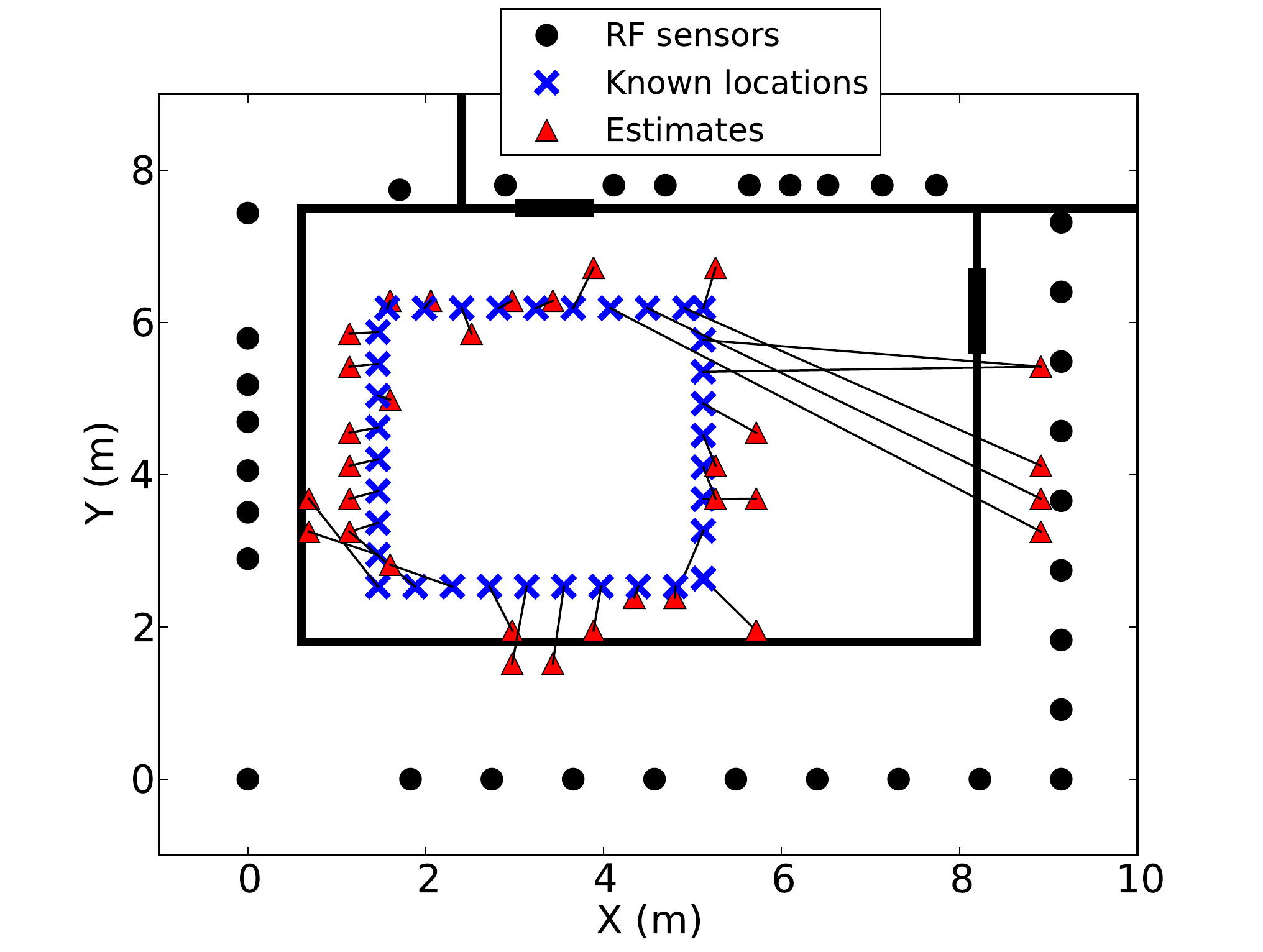}
  \caption{Estimates from VRTI.}
  \label{F:estimates_rti}
\end{figure}

\begin{figure}[htbp]
  \centering
  \includegraphics[width=3.1in]{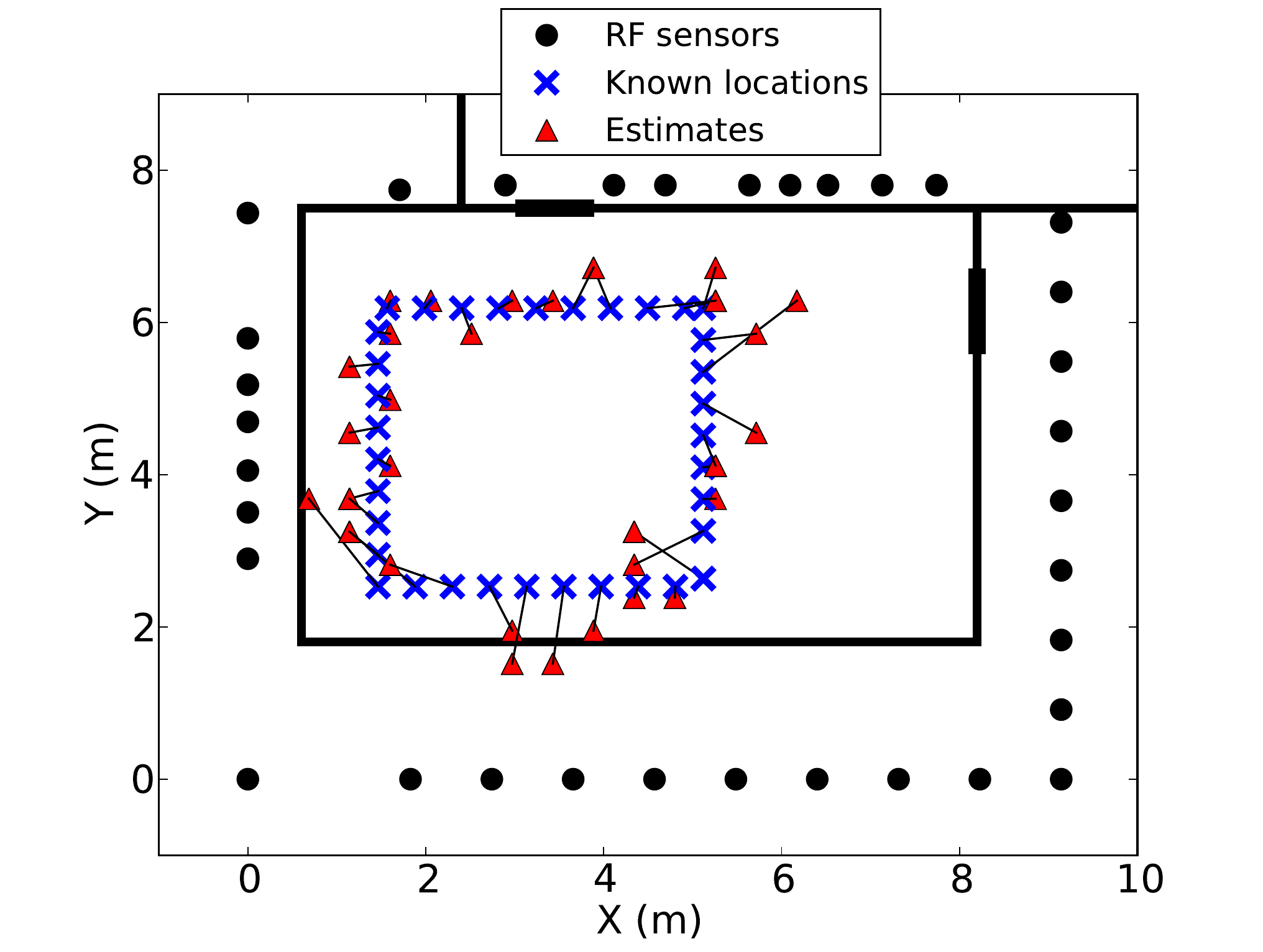}
  \caption{Estimates from SubVRT.}
  \label{F:estimates_sub}
\end{figure}

\begin{figure}[htbp]
  \centering
  \includegraphics[width=3.1in]{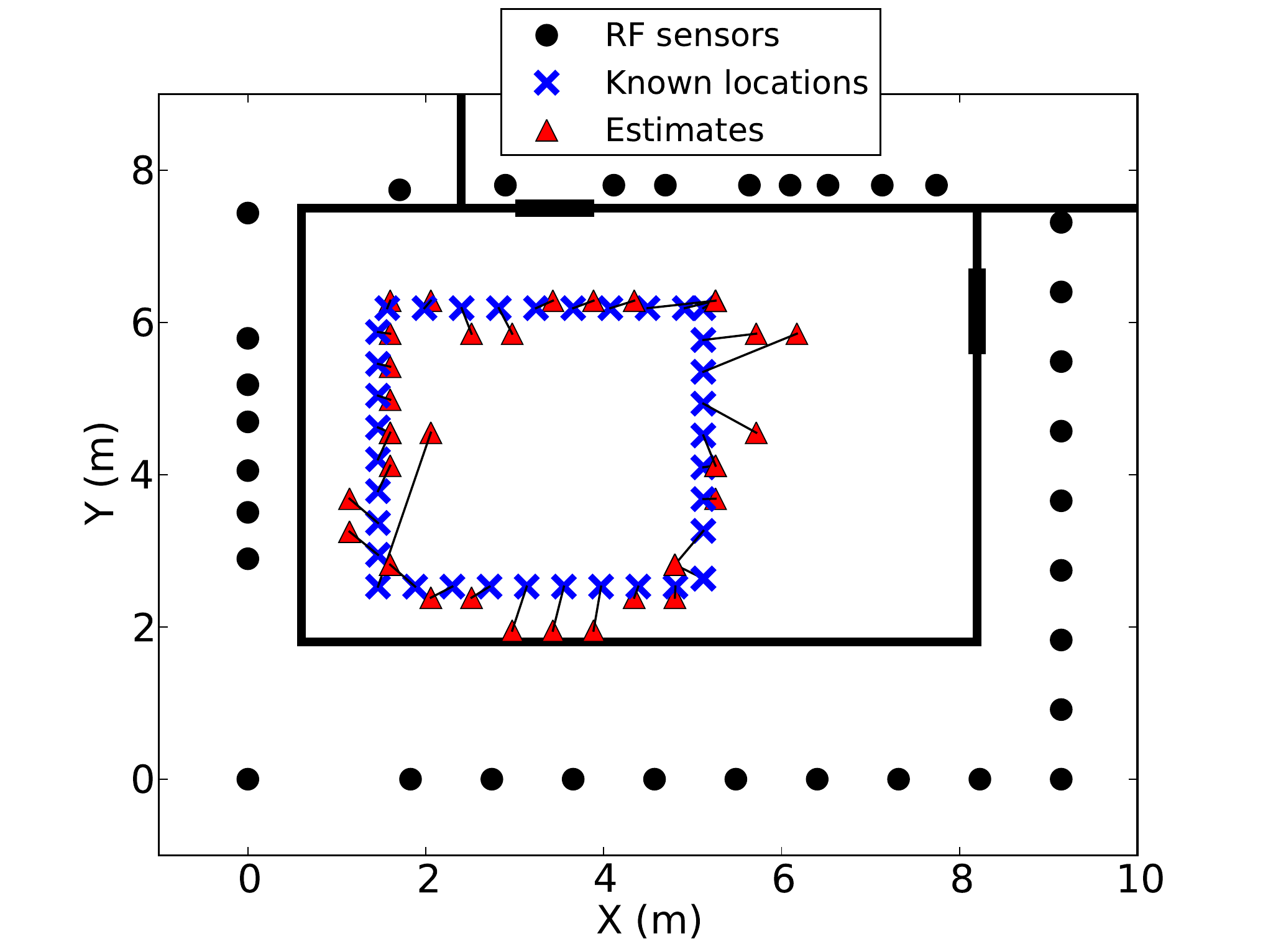}
  \caption{Estimates from LSVRT.}
  \label{F:estimates_map}
\end{figure}

The VRTI estimates of Experiment~2 are shown in Figure~\ref{F:estimates_rti}.
For clarity, we only show the actual/estimated positions when the person walks the last round of the square. 
We find that due to the impact of intrinsic motion, some estimates of VRTI are greatly biased to the right side of the experimental area (i.e., five estimates with more than 4.0~m error, as shown in Figure~\ref{F:estimates_rti}). 
However, for SubVRT and LSVRT, the impact of intrinsic motion is greatly reduced. 
As shown in Figure~\ref{F:estimates_sub} and Figure~\ref{F:estimates_map}, the estimates from SubVRT and LSVRT are more accurate than VRTI. There are no estimate errors larger than 2.0~m. 
Note that for both VRTI and SubVRT, some estimates are outside the house. The algorithms presented do not include any prior information of the house map or physical barriers which would prevent certain trajectories. 
Incorporation of prior knowledge of an indoor environment might be used to obtain better estimates, but at the expense of requiring more information to deploy the system.

\begin{figure}[htbp]
  \centering
  \includegraphics[width=3.1in]{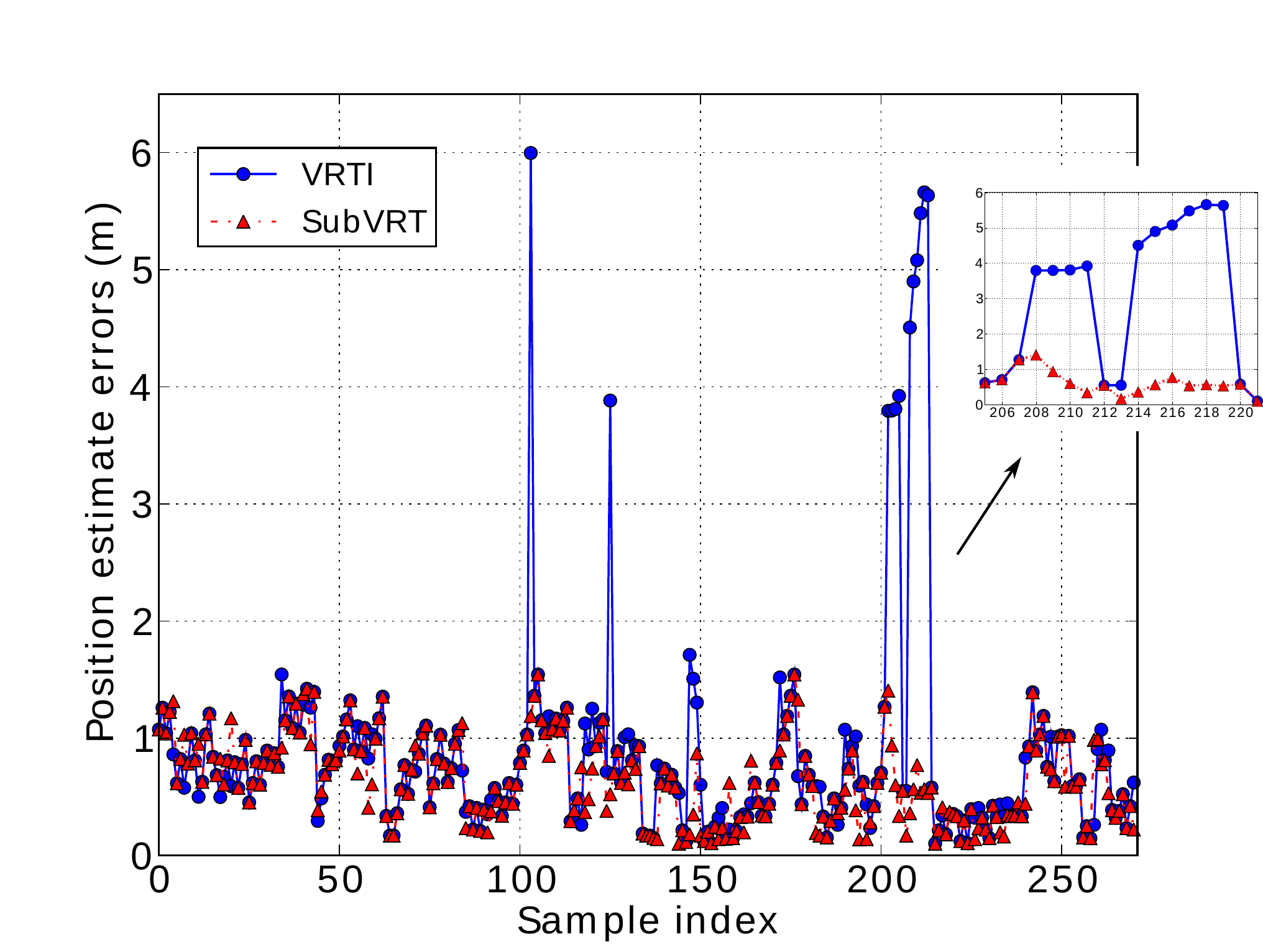}
  \caption{Estimate errors from VRTI and SubVRT.}
  \label{F:err_seq_sub}
\end{figure}

\begin{figure}[htbp]
  \centering
  \includegraphics[width=3.1in]{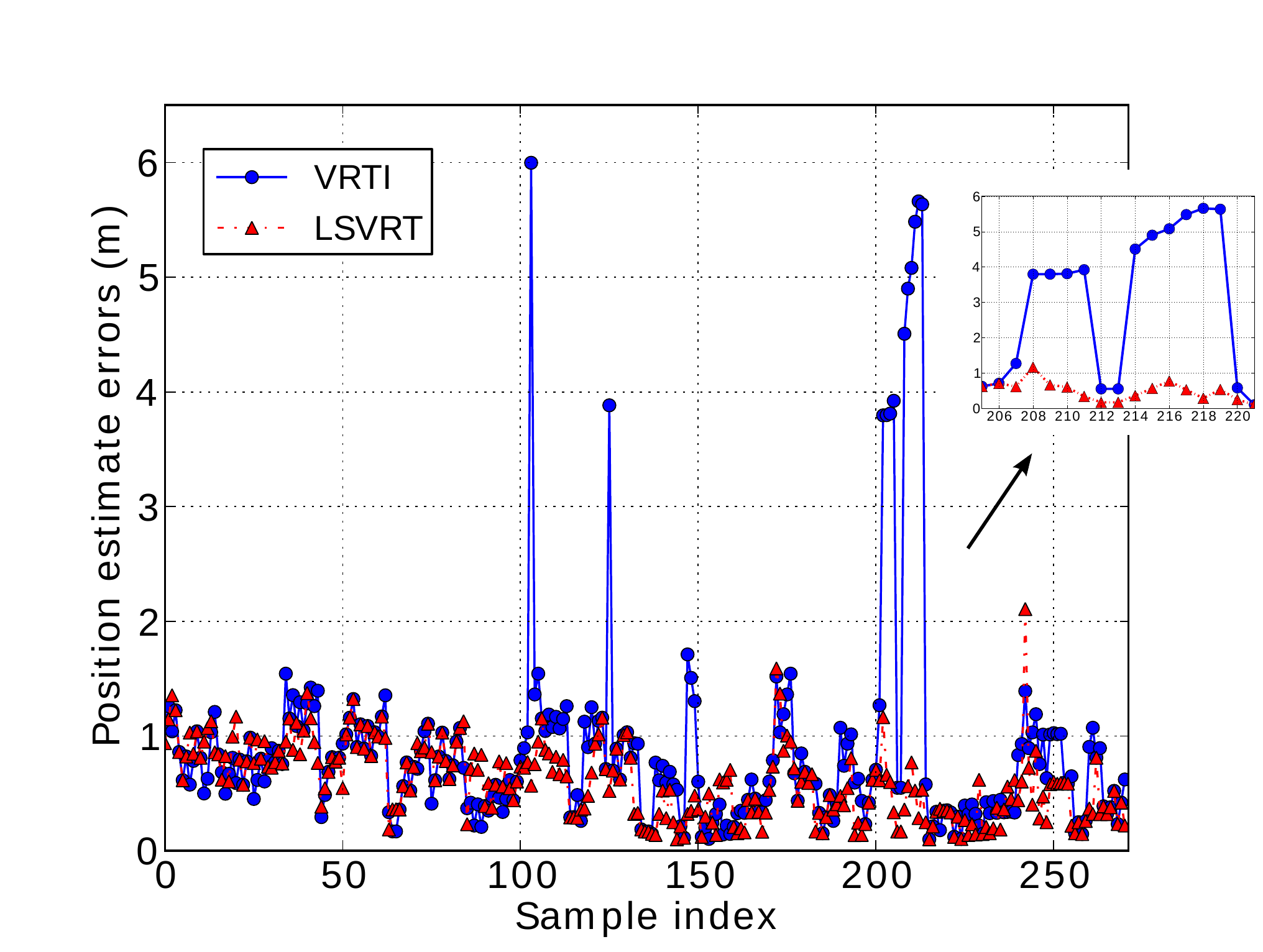}
  \caption{Estimate errors from VRTI and LSVRT.}
  \label{F:err_seq_map}
\end{figure}

Quantatively, we compare the localization errors from VRTI, SubVRT and LSVRT for the full data set.
The comparison between VRTI and SubVRT is shown in Figure~\ref{F:err_seq_sub}, and the comparison between VRTI and LSVRT is shown in Figure~\ref{F:err_seq_map}.
The localization errors from SubVRT are all below 1.8~m. For VRTI, there are several estimates with errors above 3.0~m. These large errors are due to the impact of intrinsic motion on static link measurements. 
Specifically, we compare the localization errors during a period with strong wind, from sample index 205 to 221, as shown in the inset of Figure~\ref{F:err_seq_sub}. During this period, the average localization error from VRTI is 3.0~m, while the average error from SubVRT is 0.62~m, a 79\% improvement, and for LSVRT, it is only 0.50~m, a 83\% improvement.

We also compare the RMSE of the estimates, which is defined as the square root of the average squared localization error over the course of the entire experiment. 
The RMSEs from the two experiments are summarized in Table~\ref{T:rmse}. 
For Experiment~1, the RMSE from VRTI is 0.70~m, while the RMSE from SubVRT is 0.65~m and the RMSE from LSVRT is 0.63~m. 
Since there are not much intrinsic motion in Experiment~1, the improvement in RMSE from SubVRT is 7.0\%, and the improvement from LSVRT is 9.6\%.
For Experiment~2, the RMSE from VRTI is 1.26~m, while SubVRT and LSVRT are more robust to impact of intrinsic motion. The RMSE from SubVRT is 0.74~m, a 41.3\% improvement, and the RMSE from LSVRT is 0.69~m, a 45.3\% improvement.

\begin{table}
  \centering
  \begin{tabular}{|c|c|c|c|c|c|}
    \hline       
    Methods & VRTI & \multicolumn{2}{|c|}{SubVRT} & \multicolumn{2}{|c|}{LSVRT} \\
    \hline
    Results & RMSE & RMSE & Improvement & RMSE & Improvement \\
                   \hline
     Exp. 1  & $0.70$ & $0.65$ & $7.0\%$ & $0.63$ & $9.6\%$\\
                   \hline
     Exp. 2  & $1.26$ & $0.74$ & $41.3\%$ & $0.69$ & $45.3\%$\\
                   \hline                
  \end{tabular}
  \caption{Localization RMSEs from VRTI, SubVRT and LSVRT.}
  \label{T:rmse}
\end{table}

\subsection{Discussion} \label{S:Results_dis}
The parameters that we use in VRTI, SubVRT and LSVRT are listed in Table~\ref{T:parameter}.
Here, we discuss the effects of the number of principal components $k$ on the SubVRT localization results. We also discuss the effects of the covariance matrix parameter $\sigma_{x}^{2}$ on the performance of LSVRT.

\begin{figure}[htbp]
  \centering
  \includegraphics[width=3.1in]{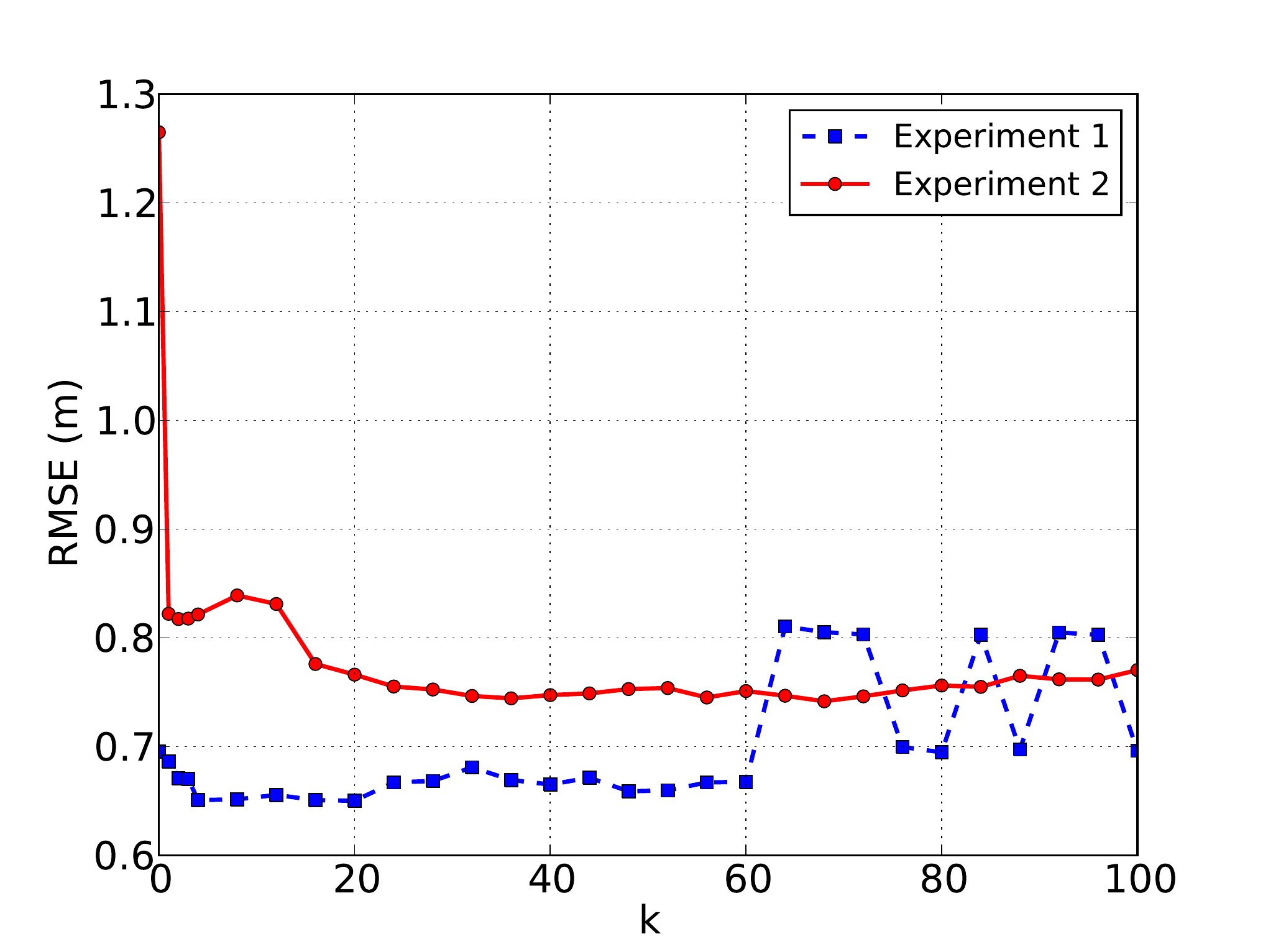}
  \caption{Localization \small{RMSE} vs. principal component number $k$.}
  \label{F:sub_k_rmse}
\end{figure}

An important parameter for SubVRT is the number of principal components used to construct the intrinsic subspace. As discussed in Section~\ref{S:Subspace}, the first $k$ components are used to calculate the projection matrix for the intrinsic subspace $\Pi_I$. If $k=0$, $\Pi_I = 0$, then $\Pi_1 = \Pi_2$, SubVRT is simplified to VRTI. 
The RMSE of SubVRT using a range of $k$ are shown in Figure~\ref{F:sub_k_rmse}. Since the first eigen-network $\mathbf{u}_{1}$ captures the strongest intrinsic signal, when $k=1$, the RMSE of Experiment~2 decreases substantially from 1.26~m to 0.82~m.
Since Experiment~1 has less intrinsic motion, the RMSE decreases from 0.70~m when $k=0$ to 0.65~m when $k=4$, a less substantial decrease.
We note that as $k$ increases, more and more information in the measurement is removed, and the RMSE stops decreasing dramatically, and even increases, at certain $k$. This is because when $k$ becomes very large, the information removed also contains a great amount of signal caused by extrinsic (human) motion. 
Thus, the performance of SubVRT could be degraded if $k$ is chosen to be too large.
The parameter $k$ is a tradeoff between removing intrinsic motion impact and keeping useful information from extrinsic motion.
For experiments without much intrinsic motion, such as Experiment~1, we choose a small $k$. However, for Experiment~2, with strong impact from intrinsic motion, we use a large $k$. As listed in Table \ref{T:parameter}, we use $k=4$ and $k=36$ for Experiment~1 and 2, respectively.

\begin{figure}[htbp]
  \centering
  \includegraphics[width=3.1in]{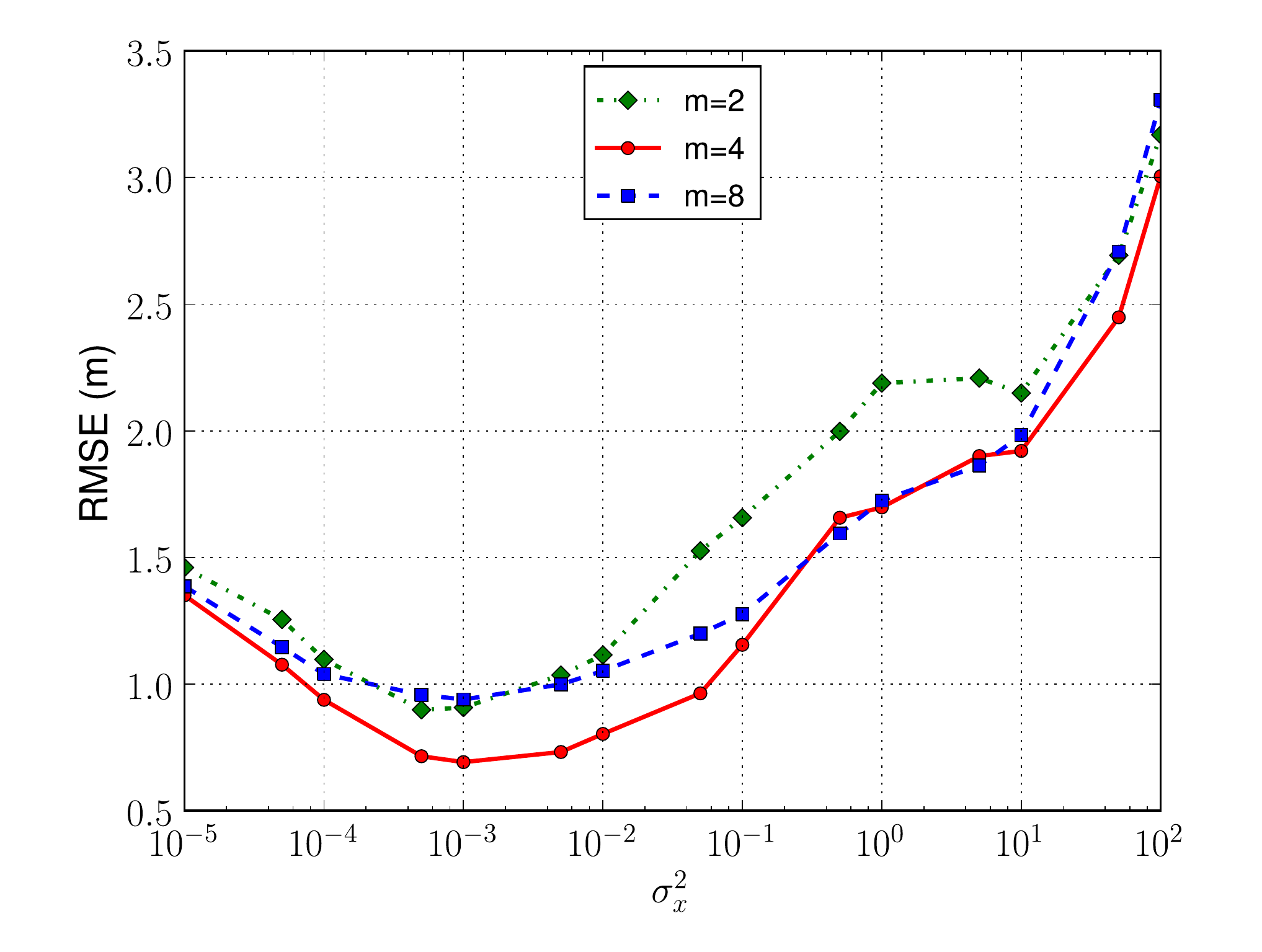}
  \caption{Localization \small{RMSE} vs. $\sigma_{x}^{2}$.}
  \label{F:rmse_sigma_x}
\end{figure}

An advantage of LSVRT over SubVRT is that LSVRT does not need to change any parameter due to changes in the environment, such as parameter $k$ in SubVRT. Thus, we only investigate parameter $\sigma_{x}^{2}$ in LSVRT, which plays the same role of the regularization parameter $\alpha$ in SubVRT. 
From Figure~\ref{F:rmse_sigma_x}, we see the RMSE from LSVRT reaches the minimum at 0.63 m, when $\sigma_{x}^{2}=0.001$ and $m=4$. 
Similar to functions of $\alpha$ shown in Figure~10 of \cite{zhao11secon}, the localization RMSEs from LSVRT are also shallow functions of $\sigma_{x}^{2}$ in the range from $10^{-4}$ to $10^{-1}$. That is, LSVRT is not very sensitive to this regularization parameter in a wide range. 

\section{Tracking} \label{S:Tracking}

In this section, we apply a Kalman filter to the localization estimates shown in Section~\ref{S:loc_results} to better estimate moving people's positions over time. Then, we compare the tracking results from VRTI with those from SubVRT and LSVRT, and show that the Kalman filter tracking results from SubVRT and LSVRT are more robust to large localization errors.

\subsection{Kalman filter}
In the state transition model of the Kalman filter, we include both position $(P_x, P_y)$ and velocity $(V_x, V_y)$ in the Cartesian coordinate system in the state vector $\mathbf{s}=[P_x, P_y, V_x, V_y]^{T}$, and the state transition model is:
\begin{equation}
  \mathbf{s} [t] = G \mathbf{s} [t-1] + \mathbf{w} [t]
\end{equation}
where $\mathbf{w}=[0,0,w_{x}, w_{y}]^{T}$ is the process noise, and $G$ is:
\begin{equation}
  G = 
  \begin{bmatrix}
    1 & 0 & 1 & 0 \\
    0 & 1 & 0 & 1 \\
    0 & 0 & 1 & 0 \\
    0 & 0 & 0 & 1
  \end{bmatrix}.
\end{equation}
The observation inputs $\mathbf{r}[t]$ of the Kalman filter are the localization estimates from VRTI, SubVRT or LSVRT at time $t$, and the observation model is:
\begin{equation}
  \mathbf{r}[t] = H \mathbf{s} [t] + \mathbf{v} [t]
\end{equation}
where $\mathbf{v}=[v_x, v_y]^{T}$ is the measurement noise, and $H$ is:
\begin{equation}
  H = 
  \begin{bmatrix}
    1 & 0 & 0 & 0 \\
    0 & 1 & 0 & 0
  \end{bmatrix}.
\end{equation}
In the Kalman filter, $v_x$ and $v_y$ are zero-mean Gaussian with variance $\sigma_{v}^{2}$, $w_x$ and $w_y$ are zero-mean Gaussian with variance $\sigma_{w}^{2}$ \cite{kay}. The parameters $\sigma_{v}^{2}$ and $\sigma_{w}^{2}$ of the measurement noise and process noise are listed in Table~\ref{T:parameter}.

\begin{table}
  \centering
  \begin{tabular}{|c|c|c|}
    \hline       
     Parameter & Value & Description \\
                   \hline          
     $\alpha$  & $100$ & Regularization parameter  \\
                   \hline    
     $m$  & $4$ & Window length to calculate variance  \\
                   \hline  
     $k$  & $4, 36$ & Numbers of principal components in Exp. 1, 2 \\
                   \hline 
     $\sigma_{x}^{2}$  & $0.001$ & Variance of human motion \\
                   \hline                    
     $\sigma_{w}^{2}$  & $2$ & Process noise parameter \\
                   \hline     
     $\sigma_{v}^{2}$  & $5$ & Measurement noise parameter \\
                   \hline
                   
  \end{tabular}
  \caption{Parameters in VRTI, SubVRT, LSVRT and Kalman filter.}
  \label{T:parameter}
\end{table}

\begin{table}
  \centering
  \begin{tabular}{|c|c|c|c|c|c|}
    \hline       
    Methods & VRTI & \multicolumn{2}{|c|}{SubVRT} & \multicolumn{2}{|c|}{LSVRT} \\
    \hline
    Results & RMSE & RMSE & Improvement & RMSE & Improvement \\
                   \hline
     Exp. 1  & $0.66$ & $0.57$ & $13.6\%$ & $0.57$ & $13.6\%$\\
                   \hline
     Exp. 2  & $1.21$ & $0.72$ & $40.5\%$ & $0.66$ & $45.5\%$\\
                   \hline                
  \end{tabular}
  \caption{Tracking RMSEs from VRTI, SubVRT and LSVRT.}
  \label{T:tracking_err}
\end{table}

\begin{figure}[htbp]
  \centering
  \includegraphics[width=3.1in]{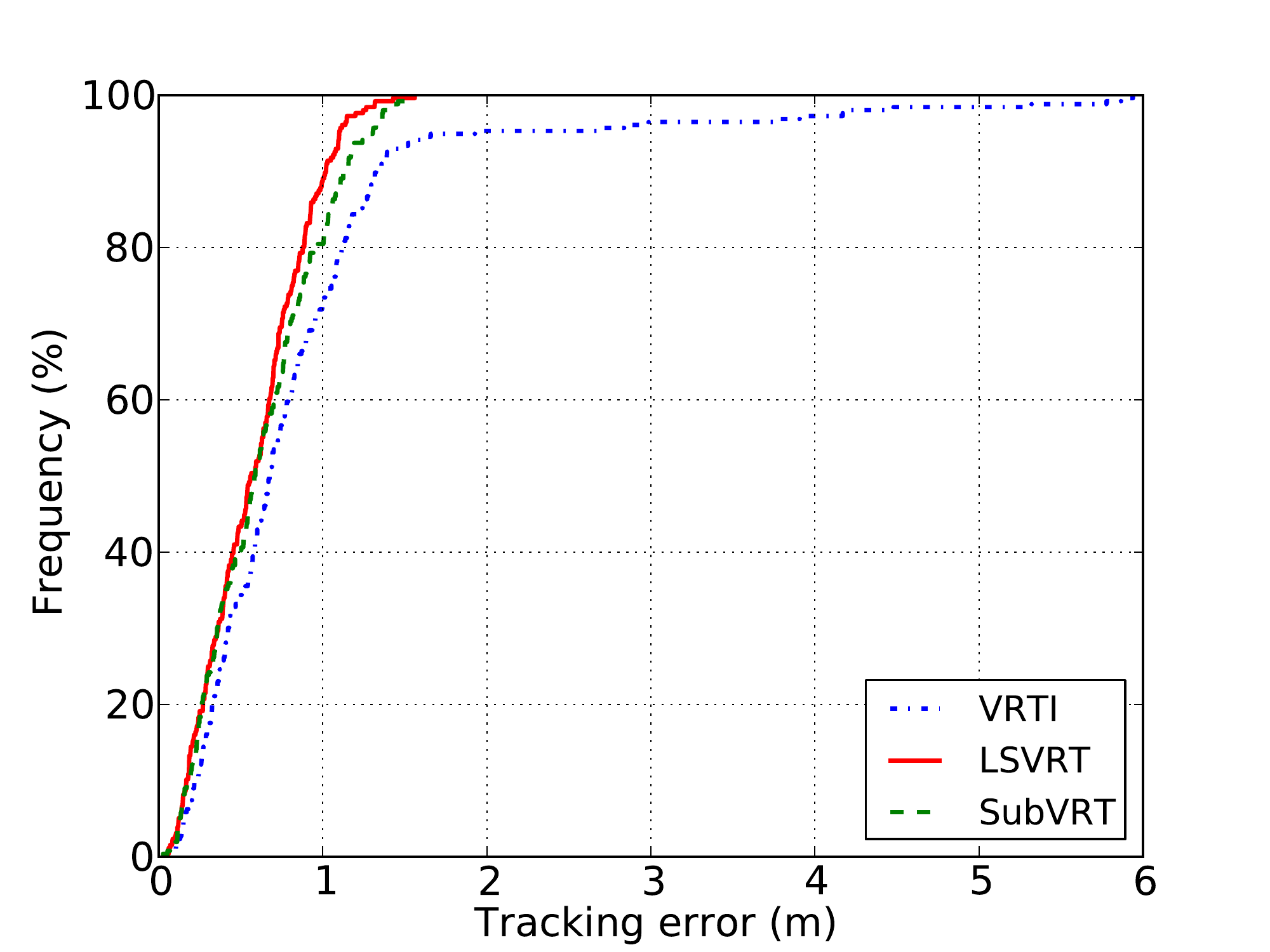}
  \caption{CDFs of tracking errors.}
  \label{F:CDF_track}
\end{figure}

\subsection{Tracking results}

We use the Kalman filter described above to track the positions of the person. 
The cumulative distribution functions (CDFs) of the tracking errors from Experiment~2 are shown in Figure~\ref{F:CDF_track}.
We see that the Kalman filter tracking results from VRTI have many more large errors than SubVRT and LSVRT. 
$97\%$ of the tracking errors from VRTI are less than 3.91~m, while $97\%$ of the tracking errors from SubVRT are less than 1.36 m, a $65.2\%$ improvement, and $97\%$ of the errors from LSVRT are less than 1.15 m, a $70.6\%$ improvement.
We use the $97^{th}$ percentile of errors to show the robustness of the tracking algorithm to large errors, and the CDFs show the tracking results from SubVRT and LSVRT are more robust to these large errors.

We also compare the RMSEs of the tracking results from VRTI, SubVRT and LSVRT, which are listed in Table~\ref{T:tracking_err}. 
For Experiment~1, the tracking RMSEs from SubVRT and LSVRT are both 0.57~m, a $13.6\%$ improvement compared to the RMSE of 0.66~m from VRTI. 
For Experiment~2, the tracking RMSE from SubVRT is reduced by $40.5\%$ to 0.72~m compared to 1.21~m RMSE from VRTI, and the RMSE from LSVRT is reduced by $45.5\%$ to 0.66~m.
We note that the tracking RMSEs from VRTI, SubVRT and LSVRT of Experiment~2 are both larger than Experiment~1 due to the impact of intrinsic motion. However, for VRTI the tracking RMSE from Experiment~2 has a $83.3\%$ increase compared to Experiment~1, while for SubVRT and LSVRT, they only increases $26.3\%$ and $15.8\%$, respectively. The tracking RMSEs from SubVRT and LSVRT are more robust to the impact of intrinsic motion.

\begin{figure}[htbp]
  \centering
  (a) \includegraphics[width=2.9in]{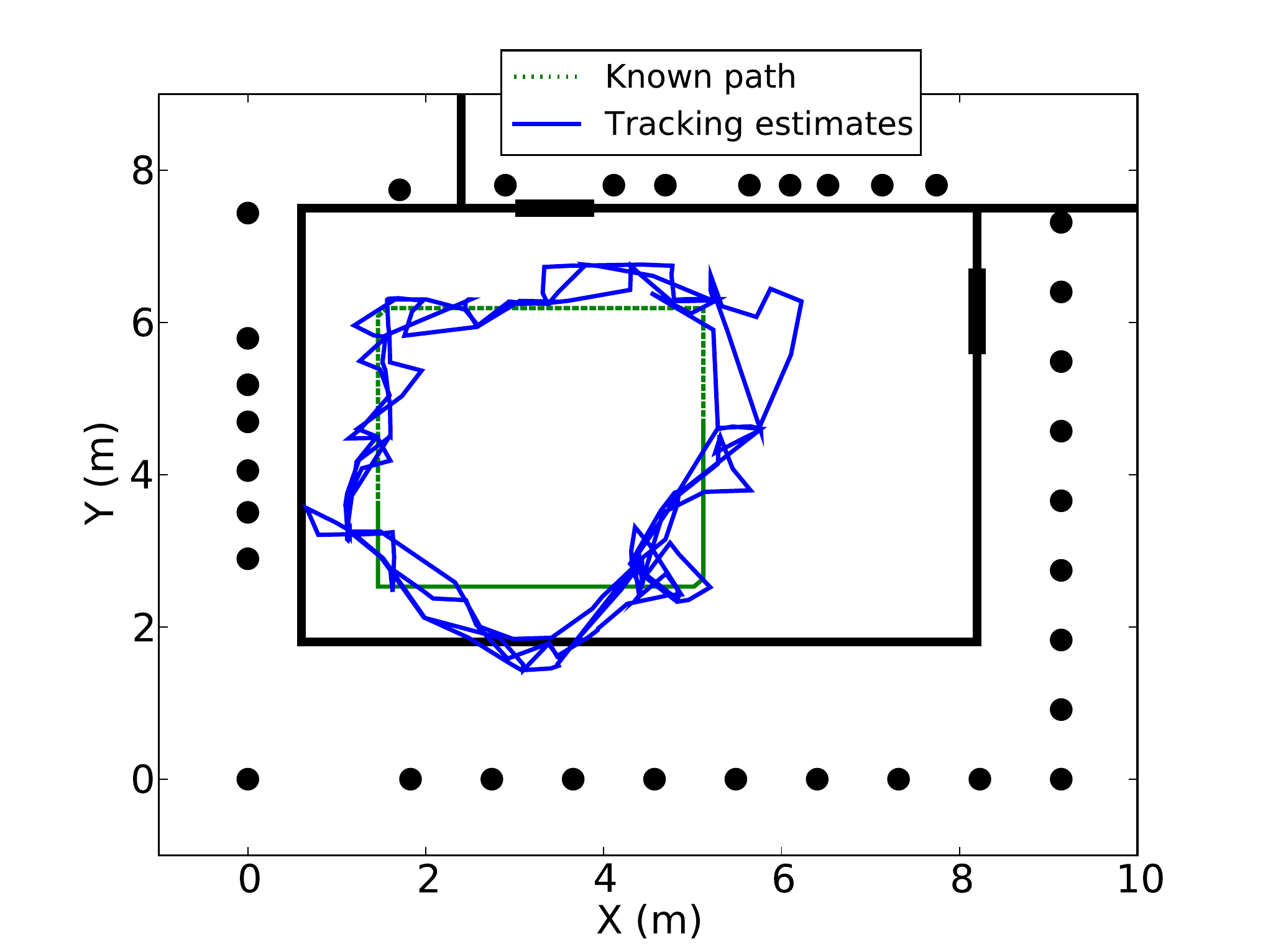} \\
  (b) \includegraphics[width=2.9in]{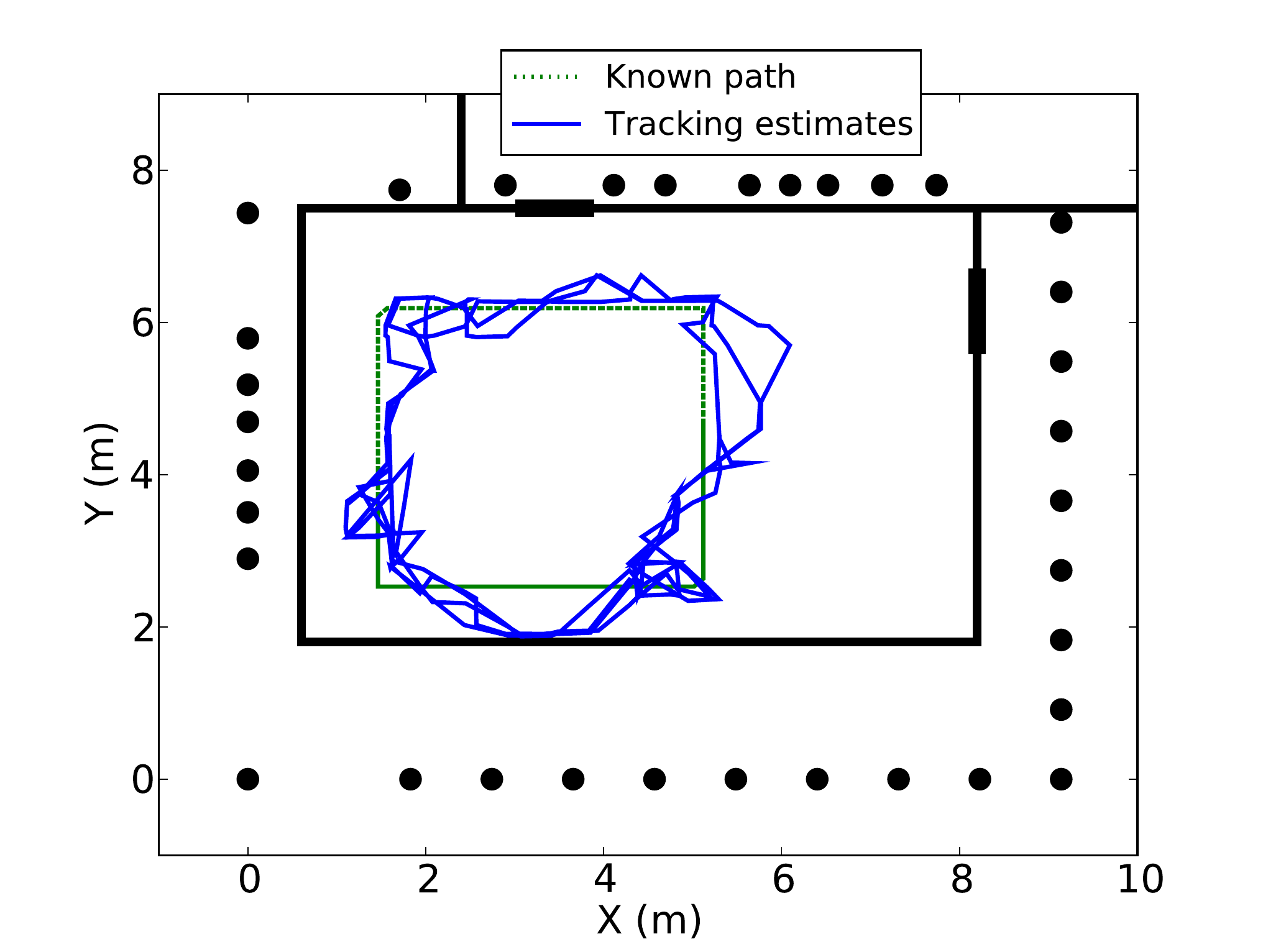}
  \caption{Kalman filter tracking results of Experiment 2 from SubVRT (a) and LSVRT (b).}
  \label{F:track_exp}
\end{figure}

Finally, the Kalman filter tracking results of Experiment~2 from SubVRT and LSVRT are shown in Figure~\ref{F:track_exp}. For Experiment~2 with significant intrinsic motion, the Kalman filter results using SubVRT and LSVRT estimates can still track a person with submeter accuracy.

\subsection{Discussion}

\begin{figure}[htbp]
  \centering
  \includegraphics[width=3.1in]{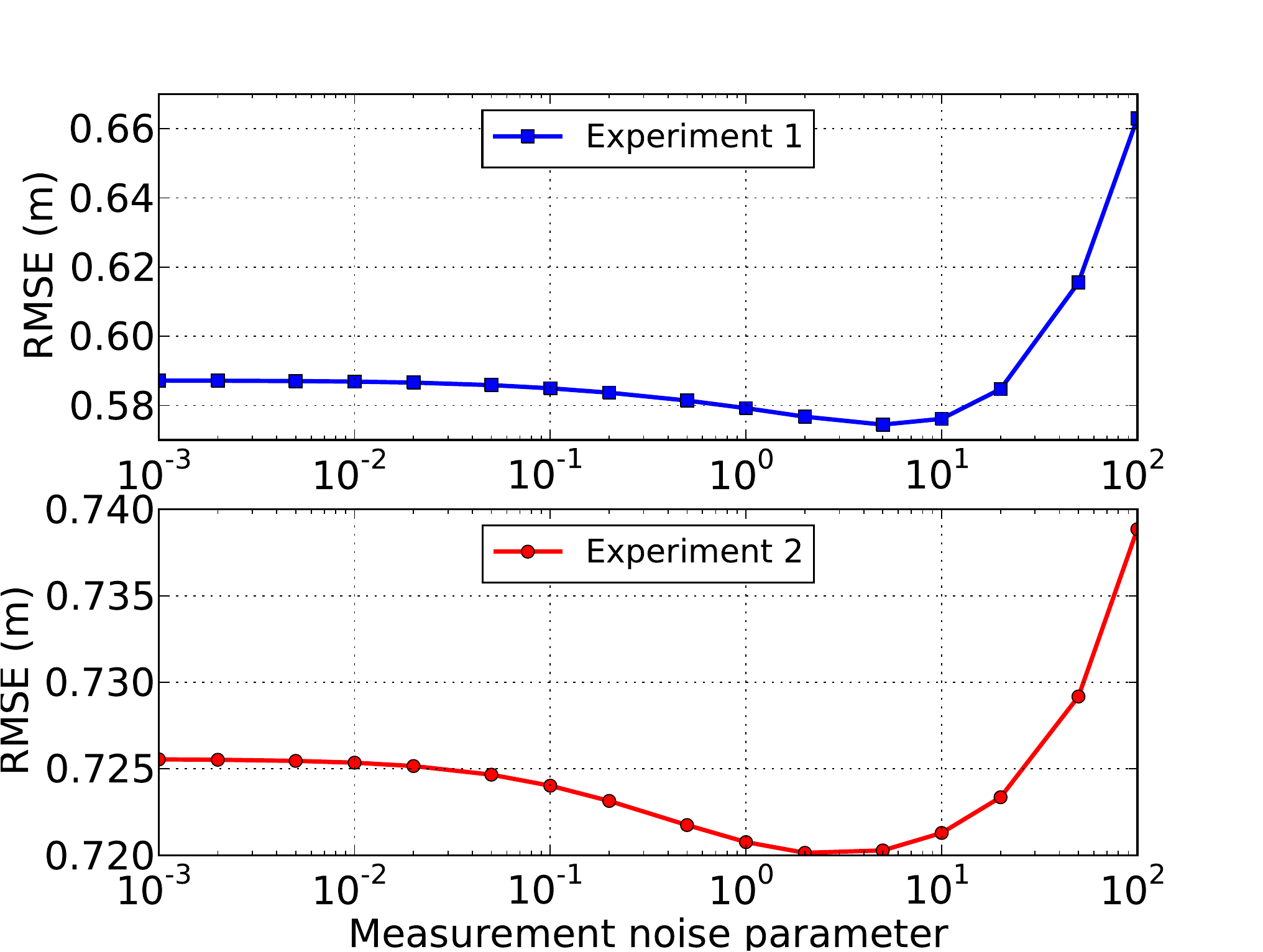}
  \caption{Tracking \small{RMSE} vs. measurement noise parameter $\sigma_{v}^{2}$.}
  \label{F:rmse_Cx}
\end{figure}

In the Kalman filter, the process noise parameter $\sigma_{w}^{2}$ should be chosen according to the dynamics of the movement. For example, for tracking vehicles, $\sigma_{w}^{2}$ should be set to a large value.
The measurement noise parameter $\sigma_{v}^{2}$ depends on how accurate the observation inputs are.
Here, we choose $\sigma_{w}^{2}$ based on the speed of moving people in typical homes, and we test the effect of using different $\sigma_{v}^{2}$ on the tracking errors.
The tracking RMSEs from SubVRT for Experiments~1 and 2 are shown as functions of $\sigma_{v}^{2}$ in Figure~\ref{F:rmse_Cx}.
If $\sigma_{v}^{2}$ is too large, the Kalman filter gives very small weights to observation inputs. On the other hand, for very small measurement noise parameter, the system dynamic model contributes little to the Kalman filter. Thus, the RMSE reaches the minimum when an appropriate balance between observation inputs and dynamic model is found. 
We also note from Figure~\ref{F:rmse_Cx} that for both Experiments, the RMSEs are shallow functions of $\sigma_{v}^{2}$ in a wide range from 0.001 to 20. That is, if we give sufficient weights to the observation inputs, which are the localization estimates from SubVRT and LSVRT, our Kalman filter tracking results are not very sensitive to the measurement noise parameter.

\section{Related work} \label{S:Related}

DFL using RF sensor networks has potential applications in surveillance for police and firefighters. Different measurements and algorithms have been proposed \cite{youssef07,zhang2007rf,viani2008object,wilson09c}. 
For RSS-based DFL, there are essentially two types of algorithms: fingerprint-based algorithms and model-based algorithms.
Like fingerprint-based real-time location service (RTLS) systems, fingerprint-based DFL methods use a database of training measurements, and estimate people's locations by comparing the measurements during the online phase with the training measurements \cite{zhang2007rf, viani2008object, seifeldin2009nuzzer}. 
Since a separate training measurement dataset is necessary, fingerprint-based DFL needs substantial calibration effort. 
As the number of people to be located increases, the training requirement increases exponentially. 
Model-based algorithms \cite{wilson09c, wilson09a, kanso09b} provide another approach. A forward model is used to relate measurements with unknown people's positions, and the localization problem can be solved as an inverse problem. 
An advantage of a model-based algorithm is that it does not need such training measurements, however, sufficient link measurements are necessary to solve the inverse problem. 
The proposed subspace decomposition and least squares methods have been applied to a model-based DFL method -- VRTI, and can significantly improve the robustness of position estimates. These methods may also be used in fingerprint-based DFL methods, but we leave this as a possible future research topic.

\section{Conclusion} \label{S:Conclusion}

In this paper, we propose to use subspace decomposition and least squares estimation to reduce noise in RSS variance-based device-free localization and tracking. 
We discuss how intrinsic motion, such as moving leaves, increase measured RSS variance in a way that is ``noise" to a DFL system. The signal caused by intrinsic motion has a spatial signature, which can be removed by the subspace decomposition method.
We apply the subspace decomposition method to VRTI, a new estimator we call SubVRT. We also propose an LSVRT estimator that directly uses the covariance matrix of the measurement to reduce the impact of intrinsic motion.
Experimental results show that SubVRT and LSVRT can reduce localization RMSE by more than $40\%$. 
We further apply a Kalman filter on SubVRT and LSVRT estimates for tracking. We find the tracking results from SubVRT and LSVRT are much more robust to large errors.

\bibliographystyle{IEEEtran}
\bibliography{overall_yang}  

\end{document}